\documentclass[12ppt,twocolumn]{emulateapj}
\usepackage{amsmath,amssymb,rotfloat}
\usepackage{graphics}
\usepackage{graphicx}
\usepackage{color}
\usepackage{ulem}

\begin{document}

\shorttitle{Magnetic field amplification and evolution in turbulent collisionless MHD}
\shortauthors{SANTOS-LIMA, DE GOUVEIA DAL PINO ET AL.}
\title{Magnetic field amplification and evolution in turbulent collisionless MHD: \\an application to the ICM}

\author{R. Santos-Lima\altaffilmark{1}, E. M. de Gouveia Dal Pino\altaffilmark{1}, G. Kowal\altaffilmark{1}, \\D. Falceta-Gon\c{c}alves\altaffilmark{2}, A. Lazarian\altaffilmark{3}, M. S. Nakwacki\altaffilmark{1}}
\altaffiltext{1}{Instituto de Astronomia, Geof\'isica e Ci\^encias Atmosf\'ericas, Universidade de S\~ao Paulo, R. do Mat\~ao, 1226, S\~ao Paulo, SP 05508-090, Brazil}
\altaffiltext{2}{Escola de Artes, Ci\^encias e Humanidades, Universidade de S\~ao Paulo, Rua Arlindo Bettio, 1000, S\~ao Paulo,
SP 03828-000, Brazil}
\altaffiltext{3}{Department of Astronomy, University of Wisconsin, Madison, WI 53706, USA}

\begin{abstract}

 The amplification and maintenance of the observed
magnetic fields in the ICM are usually attributed to the turbulent dynamo action.
This is generally derived employing a collisional MHD model. However, this assumption is poorly
justified  a priori since in the ICM the ion mean free path between collisions is of the
order of the dynamical scales, thus requiring a $collisionless$ MHD description. To deal 
with this problem on the basis of solar wind and laboratory plasmas measurements we adopt a phenomenological model of scattering of particles by magnetic perturbations arising from instabilities within collisionless plasmas. In this model,
we investigate the properties of magnetic turbulence and the  dynamo action in the ICM. Unlike collisional MHD simulations, our study uses
an anisotropic plasma pressure with respect to the direction of the local magnetic field
and this anisotropy modifies the MHD linear waves bringing the plasma within a parameter space where
collisionless instabilities should take place. However, within
the adopted model these instabilities are contained at bay through the relaxation term of the 
pressure anisotropy which simulates the
feedback of the mirror and firehose instabilities on the plasma under study. The relaxation term acts
to get the plasma distribution function consistent  with the empirical studies of collisionless plasmas. Our 
 three-dimensional numerical simulations of forced transonic turbulence motivated by modeling of the turbulent ICM are performed for different initial values of the magnetic field intensity, as well as different relaxation rates of the pressure anisotropy. We found that in the high $\beta$  plasma regime (where $\beta$ is the ratio between thermal to magnetic pressures) corresponding to the ICM conditions, a fast anisotropy relaxation rate gives results which are
similar to the collisional-MHD model as far as the statistical properties of the
turbulence are concerned. Also, the amplification of seed magnetic fields  due to the turbulent dynamo
action is similar to the collisional-MHD
model, especially in the limit of an instantaneous anisotropy relaxation. Our simulations that do not
employ the anisotropy relaxation prescription (which are more like the standard so called CGL-collisionless models)  deviate significantly  from the collisional-MHD
results, and in accordance with earlier studies, show more power at the small-scale fluctuations of both 
density and velocity representing the results of the kinetic instabilities at these scales. For
these simulations the large scale fluctuations  in the magnetic field are mostly suppressed and
the turbulent dynamo fails in amplifying seed magnetic fields and the magnetic energy saturates at values several orders of magnitude smaller than the kinetic energy. 
 
\end{abstract}


\section{Introduction}

Clusters of galaxies constitute the largest structures of the universe and most of their barionic mass is in the hot X-ray emitting gas which fills the intracluster medium (ICM). There, densities and temperatures are typically $10^{-2}-10^{-3}$ particles/cm$^{-3}$ (in the central regions) and $10^{7}-10^{8}$ K, respectively (\citealt{lagana_etal_2008}). Measurements of synchrotron emission and Faraday rotation of polarized radiation support the existence of magnetic fields in the ICM with inferred intensities of $\sim 1-10 \mu$G (\citealt{govoni_feretti_2004,ensslin_etal_2005}). At the same time, the ICM is expected to be turbulent due to the existence of numerous sources of turbulence there, the most energetic of which originate in mergers of galactic sub-clusters (see e.g., \citealt{lazarian_2006a,durret_limaneto_2008,govoni_feretti_2004,ensslin_etal_2005}).

The large scale dynamics of the magnetized plasma in the ICM, commonly described by the magnetohydrodynamic (MHD) theory, links the evolution of the observed magnetic fields and the bulk motions of the gas. In this context, one of the most important outcomes from the MHD approximation is the ability of a driven turbulent flow to amplify the magnetic fields until close equipartition between kinetic and magnetic energies (\citealt{schekochihin_2004}). That is, once a seed magnetic field is present, turbulence will stretch and fold the field lines until the magnetic forces become dynamically important. Recent collisional MHD studies (\citealt{cho_etal_2009, beresnyak_2012a}, see also Brandenburg et al. 2012, de Gouveia Dal Pino et al. 2013 for reviews) show an efficient magnetic field amplification independent of the value of the
initial seed field. This turbulent dynamo transfers about 6 percent of the energy flux of hydrodynamic cascade into magnetic energy. As the magnetic field enters into equipartition with hydrodynamic motions, the magnetic fields get correlation lengths of the order of the largest scales of the turbulence. While the origin of  seed fields is still a matter of discussion, it little depends
on the original value of the seed field and different processes, e.g. of Biermann battery, can provide the sufficient value of the initial field 
(see e.g. \citealt{lazarian_1992, grasso_rubinstein_2001, dgdp_2006, dgdp_2010}). The above turbulent dynamo scenario can
be sustaining the 
 magnetic fields in the ICM (\citealt{brandenburg_subramanian_2005, brandenburg_etal_2012, dgdp_etal_2013} and references therein). This picture is supported by MHD simulations of galaxies merger, showing the amplification of the magnetic field in the intergalactic medium (\citealt{kotarba_etal_2011}). 

However, one may wonder whether the results obtained within an MHD approximation  
can be applicable to the collisionless ICM. 
Indeed, due to the small collision frequency of the protons compared to the frequencies of the turbulent motions and to the gyrofrequency of these particles around the field lines, the typical time and distance scales involved are: (i) for the injection scales of the turbulence: $\tau_{turb} = l_{turb}/v_{turb} \sim 100$ Myr considering $l_{turb} \sim  500$ kpc and $v_{turb} \sim 10^{3}$ km s$^{-1}$ (e.g. \citealt{lazarian_2006a}); (ii) for the proton-proton (electron-electron) collision: $\tau_{pp} \sim 30$ Myr ($\tau_{ee} \sim 1$ Myr) and the mean-free path is $l_{pp} \sim 30$ kpc ($l_{ee} \sim 1$ kpc); (iii) for the proton (electrons) gyromotion: $\tau_{cp} \sim 10^{3}$ s ($\tau_{ce} \sim 1$ s) and the Larmor radius $l_{ci} \sim 10^{5}$ km ($l_{ce} \sim 10^{3}$ km). This makes the proton collision rates negligible, unabling the thermalization of the energy of their motions in the different directions, if we consider only binary collisions. As a consequence the velocity distributions of the particles parallel and perpendicular (gyromotions) to the magnetic field lines are decoupled (see  Schekochihin et al. 2005;  
\citealt{kulsrud_2005}). 

The collisionless plasmas can be very different from its collisional counterpart.  
The unavoidable occurrence of temperature anisotropy is known from kinetic theory to trigger electromagnetic instabilities (see,  for instance \citealt{kulsrud_1983}). These electromagnetic fluctuations in turn, redistribute the pitch angles of the particles, decreasing the temperature anisotropy (\citealt{gary_1993}; see also an alternative view in \citealt{schekochihin_etal_2010, rosin_etal_2011}). This instability feedback is observed in the collisionless plasma of the magnetosheath and the solar wind (\citealt{marsch_2006} and references therein), in laboratory experiments (\citealt{keiter_1999}), and kinetic simulations (e.g. \citealt{tajima_etal_1977,tanaka_1993,gary_etal_1997,gary_etal_1998,gary_etal_2000,qu_etal_2008}). On the other hand, a fluid-like model is desirable for studying the large scale plasma phenomena in the ICM, as well as the evolution of turbulence and magnetic fields there. 

Fortunately, it is possible to describe many properties of collisionless plasma still using  MHD-type models if additional
 constraints are applied. 
 The simplest \textit{collisionless} MHD-type approximation is the CGL-MHD model (\citealt{chew_etal_1956}). A modified CGL-MHD model taking into account the anisotropy constraints due to kinetic instabilities has been used for modelling the solar wind and magnetosphere in numerical simulations (\citealt{samsonov_etal_2001,samsonov_etal_2007,meng_etal_2012a, meng_etal_2012b}; see also \citealt{chandran_etal_2011} where a higher order fluid model is used). 

There have been previous important work on modeling collisionless plasma within the MHD-type approach. 
For instance,
\citet{kunz_etal_2011}, have proposed a semi-phenomenological model for heating the central regions of cold-core clusters of galaxies, which is able to counterbalance  the thermal emission losses, therefore, preventing the non observed cooling flows. This heating is originated from the conversion of turbulent to thermal energy by the micro-instabilities driven by the temperature anisotropy (see details in \citealt{kunz_etal_2011}).

In the context of rotating plasmas, \citet{sharma_etal_2006} used collisionless MHD-type simulations to study the magneto-rotational instability in accretion disks around black holes. They found that the transport of angular momentum is enhanced by the stress originated from the anisotropic pressure. 

More recently, \citet{kowal_etal_2011a} performed a study of the turbulence statistics in a collisionless MHD-type model within a double-isothermal approximation (i.e., with the temperatures parallel and perpendicular to the local magnetic field  assumed to be constant). Employing 3D simulations for a set of parameters, they demonstrated how the presence of the instabilities driven by temperature anisotropy change the structure of the dynamical variables, i.e., density, magnetic field and velocity. They reported substantial
differences for the ordinary MHD their collisionless MHD-type simulations of turbulence, for instance
substantial growth of turbulent energy at small scales.

At the same time, one may wonder whether the particles in collisionless fluids indeed behave similarly
to their description in the MHD-type simulations above. 
For instance, \cite{lazarian_beresnyak_2006} considered a fluid of magnetized cosmic rays and showed that instabilities in this fluid, such as  the
gyroresonance instability (see \citealt{gary_1993}), decrease substantially the effective mean free path of otherwise collisionless particles. They also discussed the application of the same approach to collisionless plasmas (see also \citealt{schekochihin_cowley_2006}). This approach was further developed in \cite{yan_lazarian_2011} where mean free paths of particles were calculated using fully kinetic calculations. In other words, although
the collisions of the particles between each other are unimportant, they cannot stream 
freely. Instead, they interact with the perturbations of the magnetic fields which are created by collective effects in the plasma. The very instabilities that were observed in numerical simulations (see \citealt{kowal_etal_2011a}) provide feedback on the plasma particles, decreasing their mean free path\footnote{This is a sort of self-regulation in collisionless plasmas, when  instabilities constrain their own growth through decreasing the effective mean free paths.}.

This approach stating that the actual mean free path of the particles is not, in general, determined by  Spitzer cross-sections of particles, but is due  to complex plasma feedback has been adopted in
a number of works already, as mentioned above, particularly in the solar wind and magnetosphere context 
 (Samsonov et al. 2001,
2007; Meng et al. 2012a,b;  Chandran et
al. 2011).  Also, \cite{brunetti_lazarian_2011} calculated the efficiency of 
turbulent acceleration of cosmic rays in clusters of galaxies assuming that the ICM plasma has small
effective mean free paths due to the effects discussed above. This, in agreement with observations,
 increased the efficiency of the acceleration due to the suppression of the collisionless damping of the compressible
turbulence.  \textit{In situ}  measurements of solar wind plasmas and other empirical plasma data also agree with the conclusion that collisionless plasmas do experience high rate
of collisions after all, but these collisions are not due to Coulomb interactions, but  to particles
interacting with the modes induced through instabilities. This also motivates us to explore how a
plasma similar to the observed in solar wind would behave\footnote{In terms of the simulations, one may expect that reducing the mean free path of the particles should
make a collisional MHD approach applicable to describe such environments, as in the presence of instabilities one can 
expect rather efficient scattering.}.

The present work extends the previous study of \cite{kowal_etal_2011a} and investigates  the behaviour of the turbulence and the amplification of the magnetic fields in the ICM, employing  a collisionless MHD model with pressure anisotropy which is constrained by the feedback arising from the kinetic instabilities. 
For this goal, we perform three-dimensional numerical simulations of forced turbulence in the presence of  initial magnetic fields with different intensities. In \S 2, we discuss briefly the kinetic instabilities due to pressure anisotropy and their feedback on the plasma which leads to non-linear saturation. In \S 3, we describe the one-fluid model setup for the collisionless plasma employed in this work. In \S 4, we describe our numerical experiments and results, followed by the Discussions in \S 5. In \S 6, we summarize our conclusions.

\section{MHD model for a collisionless plasma}

Measurements from weakly collisional plasmas, as those in the solar wind or the magnetosheath and laboratory experiments, as well as Particle-In-Cell (PIC) numerical simulations have
 demonstrated that the kinetic instabilities driven by pressure anisotropy are able to induce the pitch angle scattering of plasma particles, thus decreasing the resulting anisotropy. This scattering 
can be calculated from  first principles (e.g. \citealt{lazarian_beresnyak_2006, yan_lazarian_2011}), but the
corresponding calculations are rather complicated if the feedback of a few instabilities should be calculated. Therefore, in this paper we adopt an empirical approach based on the measurements above. 
The motivation for the choice of our relaxation model was also based on the fact that, regardless of the differences in  collisionless plasma regimes and the anisotropy instabilities, analytical models (\citealt{hall_1979, hall_1980, hall_1981}), quasi-linear calculations (\citealt{yoon_seough_2012, seough_yoon_2012}),  PIC simulations (\citealt{gary_etal_1997, gary_etal_1998, gary_etal_2000, nishimura_etal_2002, riquelme_etal_2012}),  laboratory experiments (\citealt{keiter_1999}), as well as all the available \textit{in situ} cosmic plasma observations (\citealt{marsch_2006} and references therein) evidence the existence of saturation of the temperature anisotropy at some level, originated from microscopic electromagnetic instabilities.

In order  
to describe the evolution of turbulence in the collisionless plasma of the ICM, we employ the one fluid CGL-MHD model
(\citealt{chew_etal_1956})
which is modified
to  take into account the anisotropy relaxation due to the feedback of kinetic instabilities. In the  next paragraphs of this section we describe our assumptions and provide further justification for them (for further discussion see also \S 5).

The derivation of the CGL-MHD equations from the Vlasov-Maxwell equations can be found, for example, in \cite{kulsrud_1983}. We can write the equations of the model  in the following conservative form:
\begin{equation*}
  \frac{\partial }{\partial t}
  \begin{bmatrix}
    \rho \\[6pt]
    \rho \mathbf{u} \\[6pt]
    \mathbf{B} \\[6pt]
    e \\[6pt]
    A (\rho^{3}/B^{3})
  \end{bmatrix}
  + \nabla \cdot
  \begin{bmatrix}
    \rho \mathbf{u} \\[6pt]
    \rho \mathbf{uu} + \Pi_{P} + \Pi_{B} \\[6pt]
    \mathbf{uB - Bu} \\[6pt]
    e \mathbf{u} + \mathbf{u} \cdot \left( \Pi_{P} + \Pi_{B} \right) \\[6pt]
    A (\rho^{3}/B^{3}) \mathbf{u}
  \end{bmatrix}
  =
\end{equation*}
\begin{equation}
  = 
  \begin{bmatrix}
    0 \\[6pt]
    \mathbf{f} \\[6pt]
    0 \\[6pt]
    \mathbf{f \cdot v}  + \dot{w} \\[6pt]
    \dot{A}_{S} (\rho^{3}/B^{3})
  \end{bmatrix}
  \rm{,}
\label{eqn:collisionless_mhd}
\end{equation}
where $\rho$, $\mathbf{u}$, $\mathbf{B}$, $p_{\perp,\parallel}$ are the primitive variables density, velocity, magnetic field, and thermal pressures perpendicular/parallel to the local magnetic field, respectively; $e = (p_{\perp} + p_{\parallel}/2 + \rho u^{2}/2 + B^{2}/8\pi)$ is total energy density, $A = p_{\perp} / p_{\parallel}$ is the anisotropy in the pressure. $\Pi_P$ and $\Pi_B$ are the gyrotropic pressure  and the magnetic stress tensors, respectively,  defined by
\begin{equation}
\Pi_{P} = p_{\perp} \mathbf{I} + (p_{\parallel} - p_{\perp}) \mathbf{bb} \rm{,}
\end{equation}
and
\begin{equation}
\Pi_{B} = (B^{2}/8 \pi) \mathbf{I} - \mathbf{BB} /4 \pi \rm{,}
\end{equation}
where $\mathbf{I}$ is the unitary dyadic tensor and $\mathbf{b} = \mathbf{B} / B$. In the source terms, $\mathbf{f}$ represents an external bulk force responsible for driving the turbulence (see details  in \S 3.1), $\dot{w}$ gives the rate of change of the internal energy $w = (p_{\perp}+p_{\parallel}/2)$ of the gas due to heat conduction and radiative cooling, and $\dot{A}_{S}$ gives the rate of change of $A$ due to microphysical processes 
(see \S 2.2). 
Before specifying the details of these source terms, we briefly present the dispersion relation of the waves and instabilities in a CGL-MHD homogeneous system.

\subsection{CGL-MHD waves and instabilities}

In the absence of the source terms in Equation~\ref{eqn:collisionless_mhd}, we recover the standard CGL-MHD model. In this model,  first obtained by \cite{chew_etal_1956},  the evolution of the pressure components is expressed  by:
\begin{equation}
\frac{d }{d t} \left( \frac{p_{\perp}}{\rho B} \right) = 0, \;\;\;\;\;
\frac{d }{d t} \left( \frac{p_{\parallel} B^{2}}{\rho^{3}} \right) = 0,
\label{eqn:cgl_closure}
\end{equation}
which is also called the double-adiabatic law.

Linear perturbation analysis of the CGL-MHD equations reveals three waves, analogous to the Alfv\'en, slow and fast magnetosonic MHD waves. These waves, however, can have imaginary frequencies for sufficiently high degrees of the pressure anisotropy. The corresponding dispersion relation can be found
in \cite{hau_wang_2007}. For convenience, we reproduce here these relations:
\begin{equation}
\left( \frac{\omega}{k} \right)^{2}_{a} = 
\left( \frac{B^{2}}{4 \pi \rho} + 
\frac{p_{\perp}}{\rho} - 
\frac{p_{\parallel}}{\rho} \right) 
\cos^{2}\theta,
\label{eqn:cgl_alfven_speed}
\end{equation}
\begin{equation}
\left( \frac{\omega}{k} \right)^{2}_{f,s} = 
\frac{b \pm \sqrt{b^{2} - 4 c}}{2},
\label{eqn:cgl_fastslow_speed}
\end{equation}
where $\cos \theta = \mathbf{k \cdot B}/B$ (being $\mathbf{k}$ the wavevector of the perturbation) and
\begin{equation*}
b = \frac{B^{2}}{4 \pi \rho} + 
\frac{2 p_{\perp}}{\rho} + 
\left(\frac{2 p_{\parallel}}{\rho} - 
\frac{p_{\perp}}{\rho} \right) 
\cos^{2}\theta ,
\end{equation*}
\begin{equation*}
c = - \cos^{2}\theta
\left[ 
\left( \frac{3 p_{\parallel}}{\rho} \right)^{2} \cos^{2}\theta - 
\frac{3 p_{\parallel}}{\rho} b + 
\left( \frac{p_{\perp}}{\rho} \right)^{2} 
\sin^{2}\theta 
\right] .
\end{equation*}

The dispersion relation for the transverse (Alfv\'en) 
mode $(\omega / k )^{2}_{a}$ coincides with that obtained from the kinetic theory (in the limit when the Larmor radius goes to zero) and does not change when heat conduction is added to the system (see \citealt{kulsrud_1983}); the criterium for the instability (named firehose instability), in terms of $A = p_{\perp} / p_{\parallel}$ and $\beta_{\parallel} = p_{\parallel} / (B^{2}/8\pi)$ is in this case
\begin{equation}
A < 1 - 2\beta_{\parallel}^{-1} .
\label{eqn:firehose_crit}
\end{equation}

However, for the compressible modes $(\omega/k)^{2}_{f,s}$ (which include the mirror unstable modes), the linear dispersion relation of the CGL-MHD equations is known to deviate from the kinetic theory. The mirror instability criterium is
\begin{equation}
A/6 > 1 + \beta_{\perp}^{-1} ,
\end{equation}
while the one derived from the kinetic theory is
\begin{equation}
A > 1 + \beta_{\perp}^{-1} ,
\label{eqn:kin_mirror_crit}
\end{equation}
where $\beta_{\perp} = p_{\perp} / (B^{2}/8\pi)$ in the last two expressions.

Taking into account the finite Larmor radius effects, \cite{meng_etal_2012a} (see also \citealt{hall_1979, hall_1980, hall_1981}) give the following expressions for the the maximum growth rate $\gamma_{max}$ (normalized by the ion gyrofrequency $\Omega_i$) of the firehose and kinetic mirror instabilities:
\begin{equation}
\frac{\gamma_{max}}{\Omega_{i}} =
\begin{cases}
\displaystyle{ \frac{1}{2} \frac{(1 - A - 2 \beta_{\parallel}^{-1})}{\sqrt{A-1/4}} } \;\;\; \text{(firehose),} \\[18pt]
\displaystyle{ \frac{4 \sqrt{2}}{3 \sqrt{5}} \sqrt{A - 1 - \beta_{\perp}^{-1}} } \;\;\; \text{(mirror),}
\end{cases}
\label{eqn:max_growth_rates}
\end{equation}
which are achieved for $k^{-1} \sim l_{ci}$, the ion Larmor radius. These expressions are valid for the case of $|A-1| \ll 1$ and $\beta_{\parallel, \perp} \gg 1$.

\subsection{Pressure anisotropy relaxation}

Following \cite{samsonov_etal_2001} and \cite{meng_etal_2012a}, whenever the plasma satisfies the firehose (Eq.~\ref{eqn:firehose_crit}) or kinetic mirror instability criteria  (Eq.~\ref{eqn:kin_mirror_crit}), 
we impose  the following pressure anisotropy relaxation condition:
\begin{equation}
\left( \frac{\partial p_{\perp}}{\partial t} \right)_{S} = - \frac{1}{2} \left( \frac{\partial p_{\parallel}}{\partial t} \right)_{S} = - \nu_{S} \left( p_{\perp} - p_{\perp}^{*} \right) ,
\label{eqn:anis_relax}
\end{equation}
where $p_{\perp}^{*}$ is the value of $p_{\perp}$ for the marginally stable state (which is obtained from the equality in Equations~\ref{eqn:firehose_crit} and~\ref{eqn:kin_mirror_crit} for each instability and with the conservation of the thermal energy $w$). 
It is not clear yet how the saturation and isotropization timescales are related to the local physical parameters. Some authors claim that the values of $\nu_{S}$ are of the order of the maximum growth rate of each instability $\gamma_{max}$, which in turn is a  fraction of the ion Larmor frequency $\gamma_{max} / \Omega_{i} \sim 10^{-2}-10^{-1}$ (see \citealt{gary_etal_1997, gary_etal_1998, gary_etal_2000}). 
 In the ICM, the frequency $\Omega_{i}$ is very large compared to the frequencies that we can resolve numerically. This implies that $\nu_{S} \rightarrow \infty$ would be a good approximation, or in other words, the relaxation to the marginal values would be instantaneous (which is similar to the ``hardwalls'' employed in \citealt{sharma_etal_2006}). However, it is not clear yet whether the extreme low density and weak magnetic fields of the ICM would result in isotropization timescales as fast as these. Therefore, we have also tested, for comparison, finite values for $\nu_{S}$ which are $\ll \Omega_{i}$ (see \S 5.1).

\subsection{Thermal relaxation}

The CGL-MHD model neglects any heat conduction or radiative cooling mechanisms which is not a realistic assumption for the ICM. In the statistically steady state of the turbulence, the constant turbulent dissipation power leads to a secular increasing of the temperature of the gas which can lead to heat conduction and radiative losses. In order to deal with these 
effects in a simplified way,
we  employ a term  $\dot{w}$ that relaxes the specific internal energy $w^{*} = (p_{\perp} + p_{\parallel}/2) / \rho$ to  the initial value $w_{0}^{*}$  at a constant rate $\nu_{th}$ (see \citealt{brandenburg_etal_1995}):
\begin{equation}
\dot{w} = - \nu_{th} ( w^{*} - w_{0}^{*} ) \rho .
\label{eqn:thermal_relax}
\end{equation}
 
Although simplistic, this approximation  is useful for two reasons: (i) it allows the system to dissipate  the turbulent power excess; and (ii) it helps to relax the local values of $w^{*}$ which  may become artificially high or low  in 
the CGL-MHD formulation without constraints on the anisotropy growth (see discussion in \S 5.2).  Combining Equations~\eqref{eqn:cgl_closure}  we find  that
\begin{equation}
  w^{*} = \left[ \left( \frac{B}{B_{0}} \right) A_{0} + \frac{1}{2} \left( \frac{B}{B_{0}} \right)^{-2} \left( \frac{\rho}{\rho_{0}} \right)^{2} \right] \frac{w_{0}^{*}}{\left( A_{0} + 1/2 \right) } ,
\label{eqn:cgl_estar}
\end{equation}
where the subscripts $0$ refer to the initial values in the Lagrangian fluid volumes.

\section{Numerical methods and setup}

\subsection{The code}

Equations~\eqref{eqn:collisionless_mhd} were evolved in a three-dimensional Cartesian box employing a modified version of the shock-capturing, second order Godunov code (\citealt{kowal_etal_2011a}). The numerical fluxes were calculated using the HLL Riemann solver, with the maximum characteristic speed evaluated from   Eqs.~\eqref{eqn:cgl_alfven_speed} and~\eqref{eqn:cgl_fastslow_speed}. For the  time integration we used the second order Runge-Kutta method (RK2).

The induction equation was integrated in its ``uncurled'' form, in an equivalent way  to the Constrained Transport method. 

To prevent negative values of the anisotropy $A$ due to precision errors during the numerical integration, we used an equivalent logarithmic formulation of the last equation in~\eqref{eqn:collisionless_mhd}, which in the absence of source terms is given by 
\begin{equation}
\frac{\partial}{\partial t} \left[ \rho \ln \left( A \rho^{2}/B^{3} \right) \right] + 
\nabla \cdot \left[ \rho \ln \left( A \rho^{2}/B^{3} \right) \mathbf{u} \right] = 0 .
\end{equation}

The pressure anisotropy relaxation was applied after each sub-time-step of the RK2 method, by transforming the conservative variables $e$ and $A$ in the primitive ones $p_{\perp}$ and $p_{\parallel}$, calculating their relaxed values through Eq.~\eqref{eqn:anis_relax} (using the same implicit method as in~\citealt{meng_etal_2012a}) and then, reconstructing back the conservative variables. 

The turbulence (represented by the source term $\mathbf{f}$ in Eq.~\ref{eqn:collisionless_mhd}) is driven 
by adding a solenoidal velocity field to the domain at the end of each time-step. This velocity field is calculated in the Fourier space with a random (but solenoidal) distribution in directions and sharply centered in a chosen value $k_{turb}$ (being the injection scale $l_{turb} = L / k_{turb}$, where $L$ is the size of the cubic domain). The forcing is approximately delta correlated in time.

The time-step constraint based on the Courant stability criterium $\delta t_{C}$ is calculated taking into account both the real and the imaginary characteristic speeds of the linear modes (Eqs.~\ref{eqn:cgl_alfven_speed} and~\ref{eqn:cgl_fastslow_speed}).

Another  time-step constraint is considered due to the thermal relaxation (Eq.~\ref{eqn:thermal_relax}). At the end of each time-step, we estimate the minimum characteristic time of the thermal relaxation $\delta t_{th}$ for the next time step as
given by
\begin{equation}
\delta t_{th} = \min \left( \frac{ w }{ \dot{w} } \right) ,
\end{equation}
where  $\dot{w}$ is the value calculated during the time-step.

The next time-step is then taken  as the minimum between $\epsilon_C \delta t_{C}$ and  $\epsilon_{th} \delta t_{th}$ where, after performing several tests, we have chosen the following 
factors $\epsilon_C = 0.3$ and $\epsilon_{th} = 0.1$.

\subsection{Reference units}

In the next sections, all the physical quantities are given in code units and can be easily converted in physical units using the reference physical quantities described bellow.

We arbitrarily choose three representative quantities from which all the other ones can be derived: a length scale $l_{*}$ (which is given by the computational box side), a density $\rho_{*}$ (given by the initial ambient density of the system), and a velocity $v_{*}$ (given by the initial sound speed in most of the models, but Model C3 for which the velocity unit is $0.3 v_{*}$;  see Table~\ref{tab:models}). For instance, with such representative quantities the physical time scale is given by the time in code units multiplied by $l_{*}/v_{*}$; the physical energy density is  obtained from the energy value in code units times $\rho_{*} v_{*}^{2}$, and so on. The magnetic field in code units is already divided by  $\sqrt{4 \pi}$, thus to obtain the magnetic field in physical units one has to multiply the  value in code units by  $v_{*} \sqrt{4 \pi \rho _{*}}$.

\subsection{Initial conditions and parametric choice}

Table~\ref{tab:models} lists the simulated models and their initial parameters. 

In Table~\ref{tab:models}, $V_{A0}$ is the Alfv\'en speed given by the initial intensity of the magnetic field directed along the $x$-axis. Initially, the gas pressure is isotropic for all the models with an isothermal sound speed $V_{S0}$. The parameter $\beta_0$ is the initial ratio between the thermal pressure and the magnetic pressure.

Turbulence was driven considering the same setup in all the models of Table~\ref{tab:models}. The injection scale is $l_{turb}=0.4$. The power of injected turbulence  $\epsilon_{turb}$ is kept constant and equal to unity. After   $t=1$  a fully  turbulent flow develops in the system with an rms velocity $v_{turb}$ close to unity. This  implies a turbulent turn-over (or cascading) time $t_{turb} \approx 0.4$.

Models A, B and C in Table 1 are collisionless MHD models with initial moderate, strong, and very small (seed)  magnetic fields, respectively. For models A and C the injected turbulence is initially super-Alfv\'enic, while for models B it is sub-Alfv\'enic.  

Amhd, Bmhd, and Cmhd correspond to collisional MHD models, i.e., have  no anisotropy in pressure. The set of equations describing these models is identical to those in Eq.~\eqref{eqn:collisionless_mhd}, but dropping the equation for the evolution of the anisotropy  $A$ and replacing the thermal energy by $w=3p/2$ (which corresponds to a politropic gas index  $5/3$). Their corresponding dispersion relations are those from the usual collisional MHD approach (rather than Equations~\ref{eqn:cgl_alfven_speed} and~\ref{eqn:cgl_fastslow_speed}).

We notice that the turbulence in the ICM is expected to be trans- or even subsonic, and the plasma beta is expect to be high ($\beta \sim 200$). Therefore, models A are possibly more representative of the typical conditions in the ICM.

In the following section we will start by describing the results for models A and B which have initial finite magnetic fields and therefore, reach a nearly steady state turbulent regime relatively rapidly after the injection of turbulence. Then, we will describe models C which start with seed magnetic fields and therefore, undergo a dynamo amplification of  field due to the turbulence and take much longer to reach a nearly steady state.  

\begin{table}[h]
\caption{Parameters of the simulated models}
\centering
\begin{tabular}{l | c c c c c c c}
\hline \hline
Name	&	$\nu_{S}$	&	$\nu_{th}$	&	$V_{A0}$	&	$V_{S0}$	&	$\beta_0$	&	$t_{f}$	&	Resolution \\
[0.5ex]
\hline
A1	&	$\infty$	&	$5$	&	$0.3$	&	$1$	&	$200$	&	$5$	&	$512^{3}$ \\

A2	&	$0$	&	$5$	&	$0.3$	&	$1$	&	$200$	&	$2$	&	$512^{3}$ \\

A3	&	$10^2$	&	$5$	&	$0.3$	&	$1$	&	$200$	&	$5$	&	$512^{3}$ \\

A4	&	$10^3$	&	$5$	&	$0.3$	&	$1$	&	$200$	&	$5$	&	$512^{3}$ \\

A5	&	$\infty$	&	$0.5$	&	$0.3$	&	$1$	&	$200$	&	$5$	&	$512^{3}$ \\

A6	&	$\infty$	&	$50$	&	$0.3$	&	$1$	&	$200$	&	$5$	&	$512^{3}$ \\

Amhd	&	-	&	$5$	&	$0.3$	&	$1$	&	$200$	&	$5$	&	$512^{3}$ \\

\hline
B1	&	$\infty$	&	$5$	&	$3.0$	&	$1$	&	$0.2$	&	$5$	&	$512^{3}$ \\

Bmhd	&	-	&	$5$	&	$3.0$	&	$1$	&	$0.2$	&	$5$	&	$512^{3}$ \\

\hline
C1	&	$\infty$	&	$5$	&	$10^{-3}$	&	$1$	&	$2 \times 10^{6}$	&	$40$	&	$256^{3}$ \\

C2	&	$0$	&	$5$	&	$10^{-3}$	&	$1$	&	$2 \times 10^{6}$	&	$40$	&	$256^{3}$ \\

C3	&	$0$	&	$5$	&	$10^{-3}$	&	$0.3$	&	$2 \times 10^{5}$	&	$40$	&	$256^{3}$ \\

C4	&	$10^2$	&	$5$	&	$10^{-3}$	&	$1$	&	$2 \times 10^{6}$	&	$40$	&	$256^{3}$ \\

Cmhd	&	-	&	$5$	&	$10^{-3}$	&	$1$	&	$2 \times 10^{6}$	&	$40$	&	$256^{3}$ \\
[1ex]
\hline
\end{tabular}
\label{tab:models}
\end{table}

\section{Results}

\begin{figure*}[h]
\centering
\begin{tabular}{c c}
\input{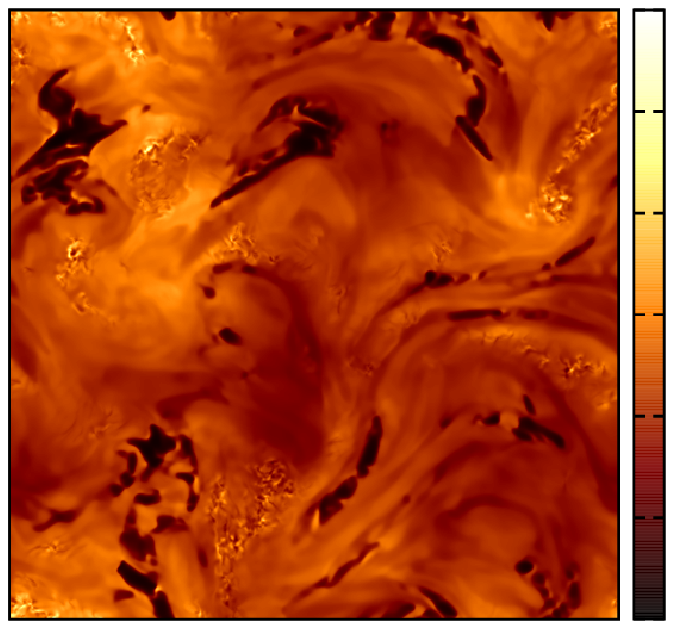} &
\input{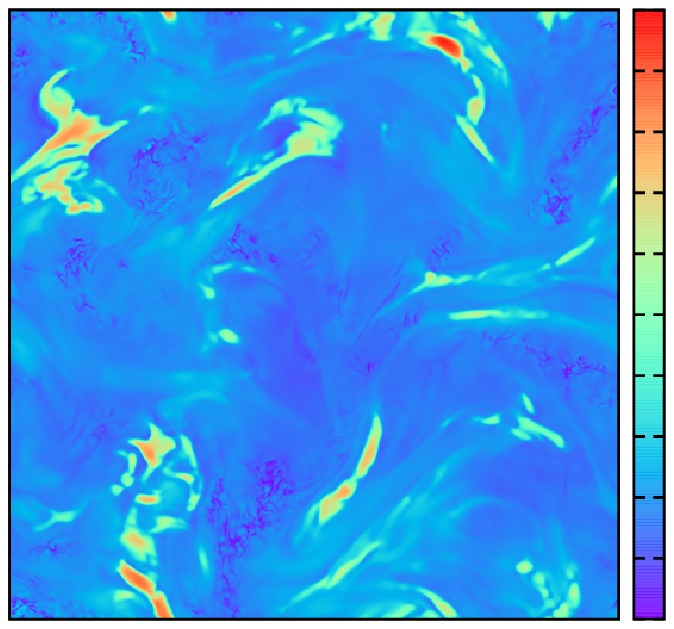} \\
\input{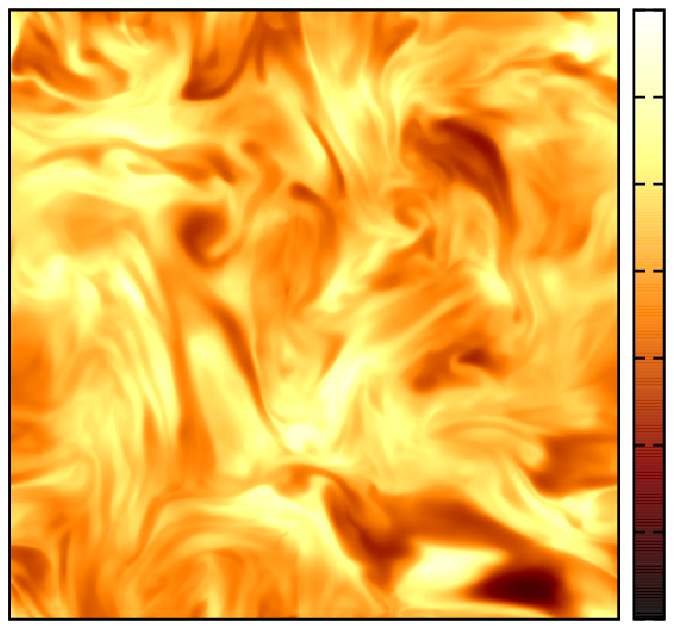} &
\input{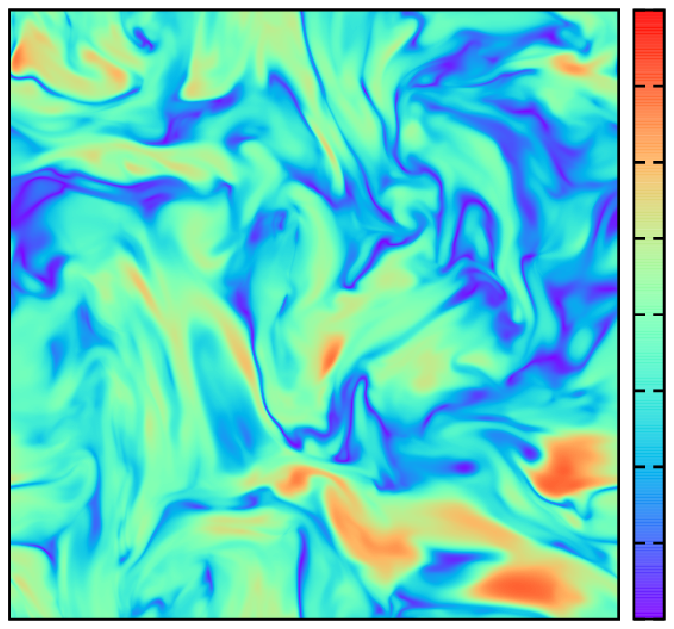} \\
\input{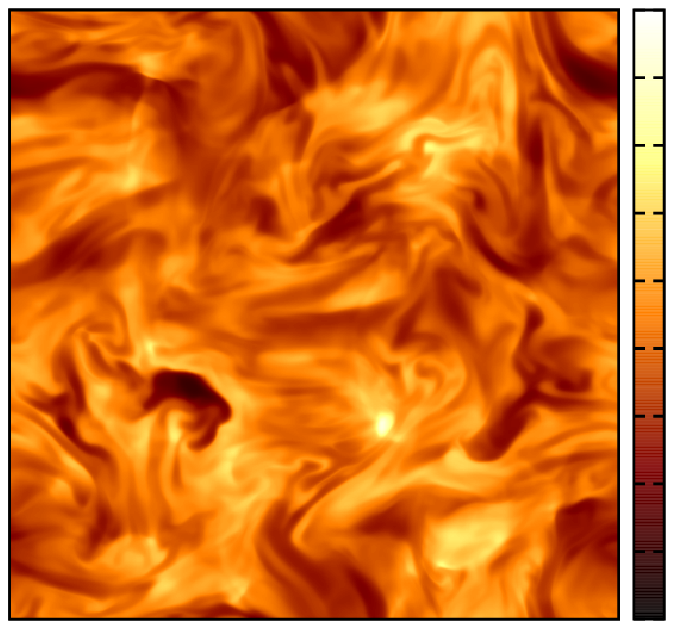} &
\input{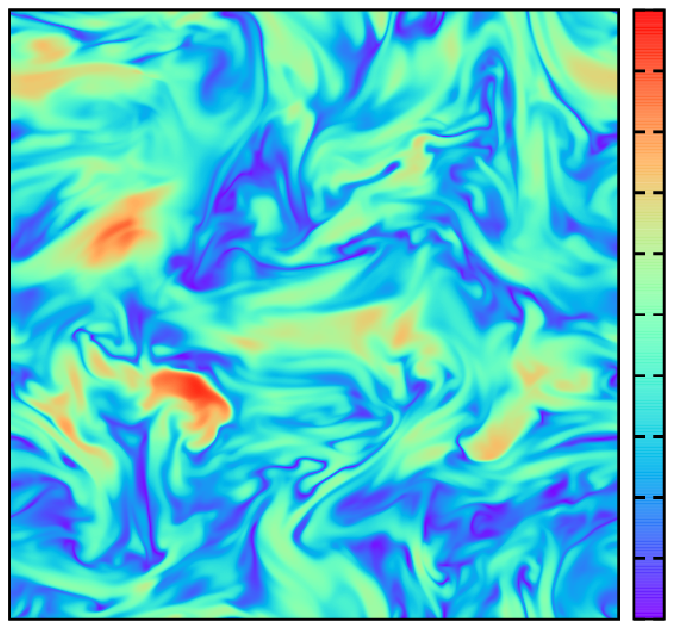}
\end{tabular}
\caption{Central XY plane of the cubic domain showing the density (left column) and the magnetic intensity (right column) distributions for  models of Table~\ref{tab:models} with initial moderate magnetic field ($\beta_0 = 200$) and  different values of the anisotropy relaxation rate $\nu_{S}$, at $t=t_f$. Top row: model A2 (with $\nu_S=0$, corresponding to the standard CGL model with no constraint on anisotropy growth); middle row: model A1 ($\nu_S = \infty$,  corresponding to instantaneous anisotropy relaxation to the marginal stability condition); bottom row: model Amhd (collisional MHD with no anisotropy). The  remaining initial conditions are all the same for the three models (see Table~\ref{tab:models}).}
\label{fig:maps_dens}
\end{figure*}

Figure~\ref{fig:maps_dens} depicts the density (left column) and the magnetic intensity (right column) distribution maps of the central slices for  collisionless models with moderate initial magnetic fields A2 (top row), A1 (middle row), and Amhd (bottom row). All  these models have  $\beta_0= 200$ and the same initial conditions, except for the anisotropy relaxation rate $\nu_{S}$.
%

In the A2 model there is no constraint on the growth of the pressure anisotropy ($\nu_S = 0$). In this case, the
kinetic instabilities  that develop due to the anisotropic pressure are very strong at the smallest scales.
This makes 
 the density (and the magnetic field intensity) distribution in Figure~\ref{fig:maps_dens} more ``wrinkled'' than in the standard (collisional) MHD case.
On the other hand, in the A1 model where the isotropization of the thermal pressure due to the back reaction of the same  kinetic instabilities is allowed to occur above a threshold, the developed density (and magnetic field intensity) structures are larger and more similar to those of the collisional MHD turbulent model Amhd.

In order to  better quantify and understand the results evidenced by Figure~\ref{fig:maps_dens} regarding  the collisionless models without and with anisotropy growth constraints, in the next paragraphs of this Section we will present a statistical analysis of the physical variables of these turbulent models after they reach a steady state.

For  models A1 to Bmhd in Table 1, the statistical analyses were performed by averaging data  from snapshots taken every $\Delta t = 1$, from $t=2$ until the final time step $t_{f}$ indicated in Table~\ref{tab:models}. For the models with initial seed fields, C1 to Cmhd, the statistical analysis considered snapshots from $t=t_f-10$ until $t_f$.

Averages and standard deviation of important physical quantities that will be discussed below are presented in Tables~\ref{tab:statistics_A},~\ref{tab:statistics_B}, and~\ref{tab:statistics_C} (in the Appendix).

\subsection{The role of the anisotropy and instabilities}

The injected turbulence produces shear and compression in the gas and in the magnetic field. Under the collisionless approximation, according to Eqs.~\eqref{eqn:cgl_closure}  $A \propto B^{3} / \rho^{2}$, therefore,  one should  expect that compressions
 along the magnetic field lines,  which keep $B$ constant but make $\rho$ to  increase,  cause a  decrease of  $A$, while compressions or shear perpendicular to the magnetic field lines, which make $B$ to increase but  keep either $B/\rho$  or $\rho$ constant, cause an increase of $A$. Therefore, even starting with $A=1$, parcels of the gas with $A \neq 1$ will naturally develop. Inside these parcels, kinetic instabilities can be triggered which in turn will inhibit the growth of the anisotropy.

\begin{figure*}
\centering
\begin{tabular}{c c}
\input{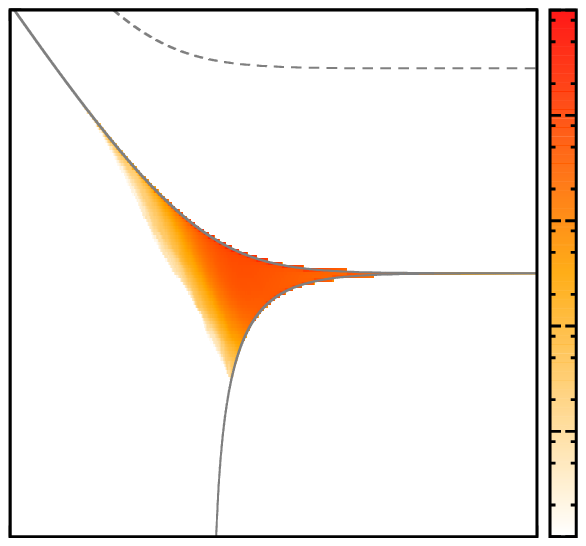} &
\input{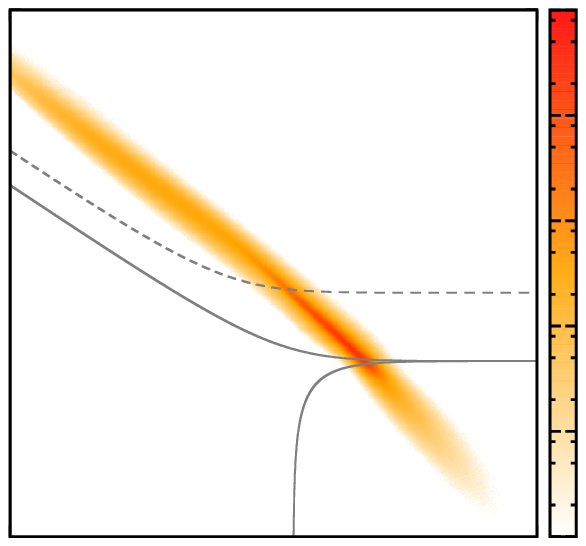} \\
\input{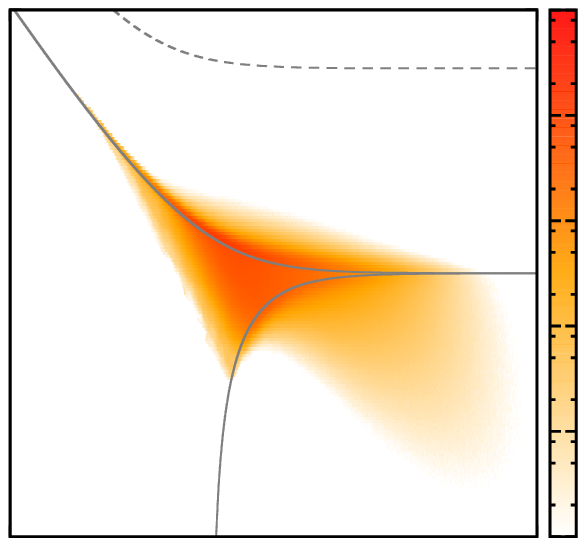} &
\input{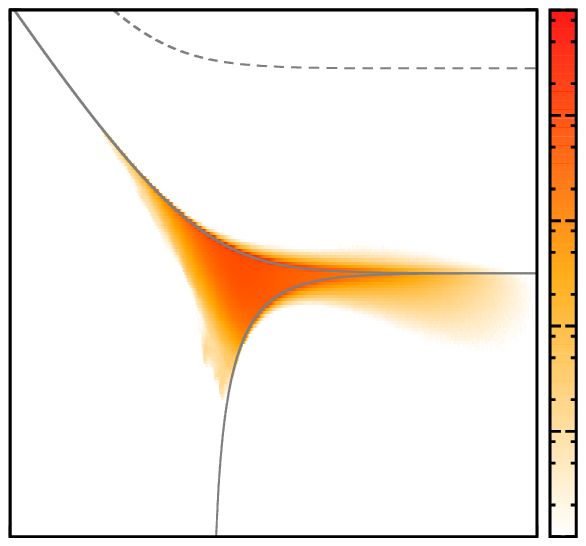}
\end{tabular}
\caption{The panels show two-dimensional normalized histograms of $A = p_{\perp}/p_{\parallel}$ versus $\beta_{\parallel} = p_{\parallel} / (B^{2}/8\pi)$  for models starting with moderate magnetic fields (models A with $\beta_0=200$) (see Table~\ref{tab:models}).
The histograms were calculated considering   snapshots every $\Delta t = 1$, from $t=2$ until the final time step $t_{f}$ indicated in Table~\ref{tab:models} for each model. The continuous gray lines represent the 
thresholds for the linear  firehose ($A = 1 - 2\beta_{\parallel}^{-1}$, lower curve) and mirror ($A = 1 + \beta_{\perp}^{-1}$, upper curve) instabilities, obtained from the kinetic theory. The dashed gray line corresponds to the linear mirror instability threshold obtained from the CGL-MHD approximation ($A/6 = 1 + \beta_{\perp}^{-1}$).}
\label{fig:beta_vs_a}
\end{figure*}

Figure~\ref{fig:beta_vs_a} presents the distribution  of the anisotropy $A$ as a function of $\beta_{\parallel}$ for the models with moderate initial magnetic field A1, A2, A3, and A4 of Table~\ref{tab:models}. 
Model A2 (which has no constraints on anisotropy growth)  has an $A$  distribution that nearly follows a line with negative inclination in the log-log diagram. This is consistent with the derived $A$ dependence in  the CGL models given by $A \propto (\rho/B) \beta_{\parallel}^{-1}$
(when the initial conditions are homogeneous; see Eqs.~\ref{eqn:cgl_closure}). 
This model attains
 values of $A$ spanning  several orders of magnitude (from $10^{-2}$ to $10^{3}$). 

Model A1, on the other hand, which has  the anisotropy bounded by the instabilities threshold values (with infinite anisotropy relaxation rate $\nu_S$), keeps $A$ close to unity, 
varying by less than one order of magnitude.

The bottom panels of Figure~\ref{fig:beta_vs_a} show the distribution of  $A$ 
for the A3 and A4  models which have bounded anisotropy with  finite anisotropy relaxation rates $\nu_S$ (see Table~\ref{tab:models}).
We see that in these cases, a fraction of the gas has $A$  values out of the stable zone. The model with smaller 
anisotropy relaxation rate (model A3) obviously  presents a larger fraction of  gas inside the unstable zones. We also note that the  higher the  value of $\beta_{\parallel}$,  the larger the linear growth rate of the instabilities and  more gas is inside the unstable regions with $A<1$. This  is consistent with the CGL trend for which $A \propto \beta_{\parallel}^{-1}$.

\begin{figure}
\centering
\begin{tabular}{c}
\input{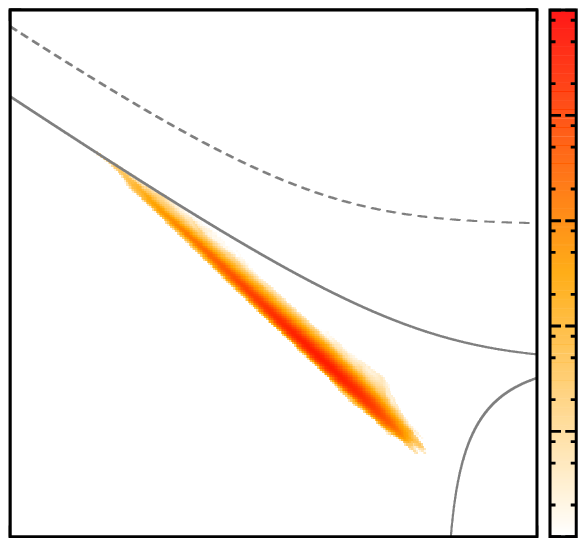}
\end{tabular}
\caption{The same as in  Figure~\ref{fig:beta_vs_a}, but  for  model B1 starting with strong magnetic field (small $\beta_0=0.2$, see Table~\ref{tab:models}).}
\label{fig:beta_vs_a_B}
\end{figure}

Figure~\ref{fig:beta_vs_a_B} shows the distribution of $A$ versus $\beta_{\parallel}$ for the model B1 with strong initial magnetic field (small $\beta_0= 0.2$). We see that in this regime, B1 model 
has an $A$ distribution 
inside the stable zone.

\begin{figure*}
\centering
\begin{tabular}{c c}
\input{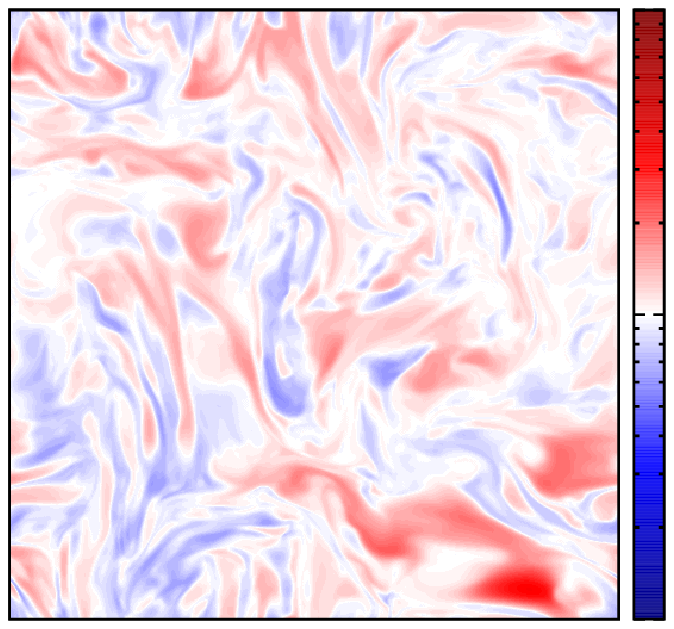} &
\input{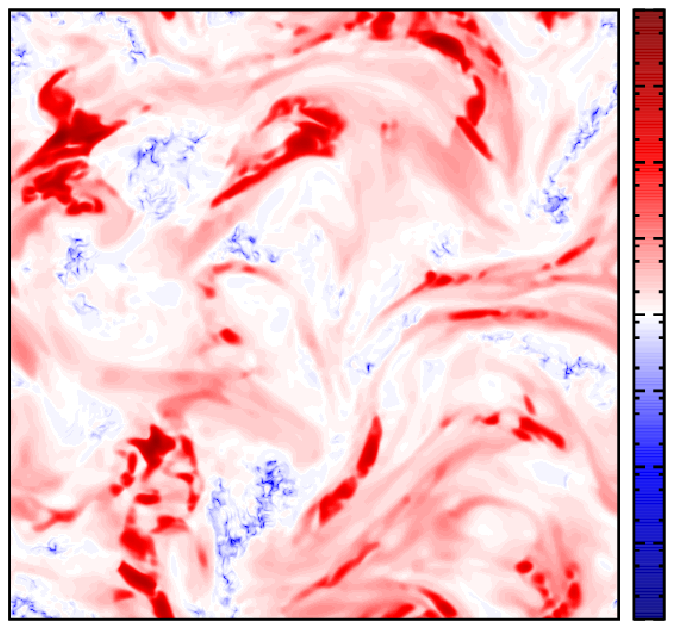} \\
\input{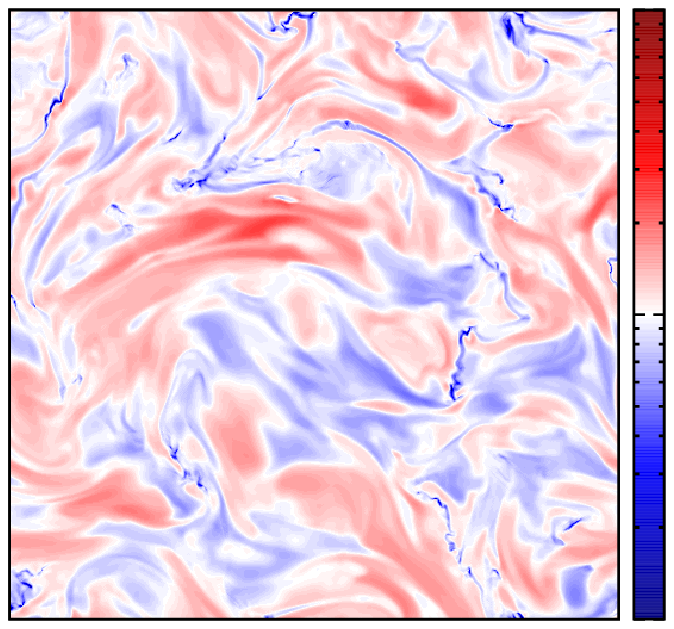} &
\input{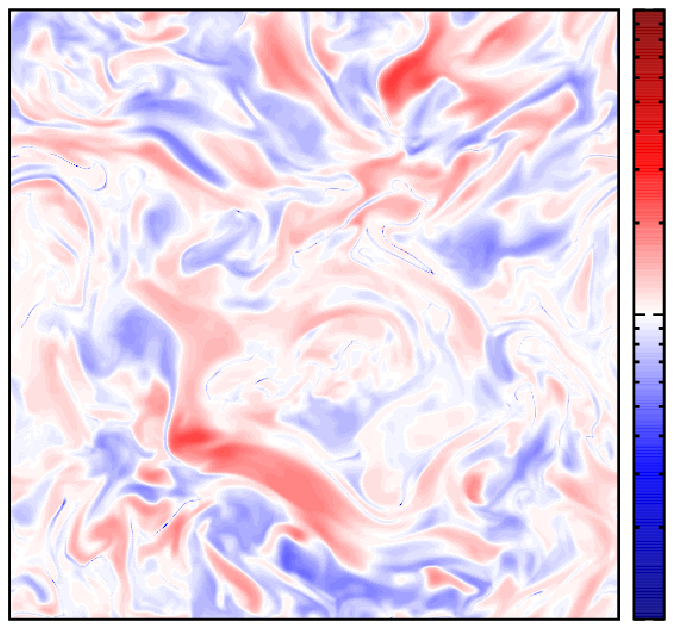} \\
\input{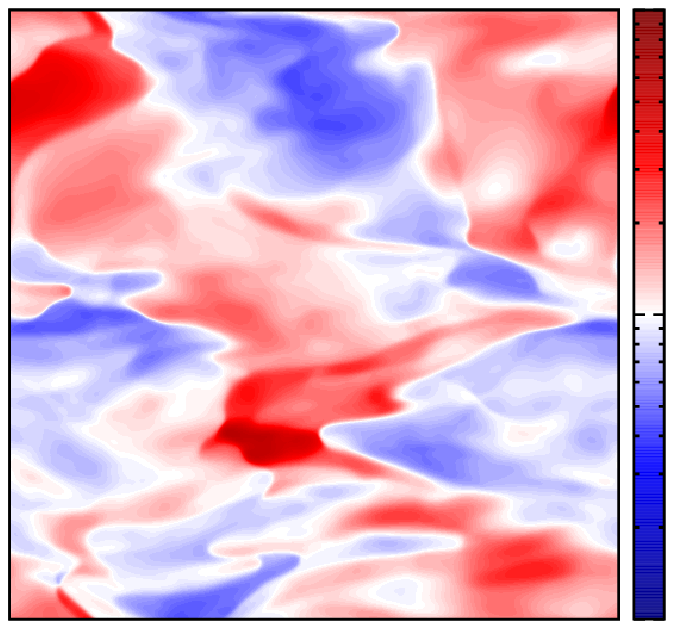}
\end{tabular}
\caption{Maps of the anisotropy $A = p_{\perp}/p_{\parallel}$ distribution at the central slice in the XY plane at the the final time $t_{f}$ for a few models A and B  of Table~\ref{tab:models}. Top and middle rows: models with moderate magnetic field (starting with $\beta_0=200$). Bottom: model with strong magnetic field (starting with $\beta_0=0.2$).  }
\label{fig:maps_anis}
\end{figure*}

The spatial anisotropy distribution is illustrated in Figure~\ref{fig:maps_anis} in two-dimensional maps  that depict central slices of $A$ in the XY-plane at the final time step for models A1, A2, A3, A4 and B1. 
For the CGL model with moderate magnetic field  ($\beta_0 = 200$), model A2, the  $A$ structures  are thin and  elongated.
These small scale structures probably arise from the fast fluctuations driven by the kinetic instabilities (see Figure~\ref{fig:beta_vs_a}).
For the model with strong magnetic field (small $\beta_0$), B1 model, the $A$ structures  are smoother.
They are originated by small amplitude magnetic fluctuations (Alfv\'en waves) and also  compression modes  at the large scales. 
The map of  model A1 also shows thin and elongated structures, but with lengths of the order of the turbulence scale. 

\begin{figure*}
\centering
\begin{tabular}{c c}
\input{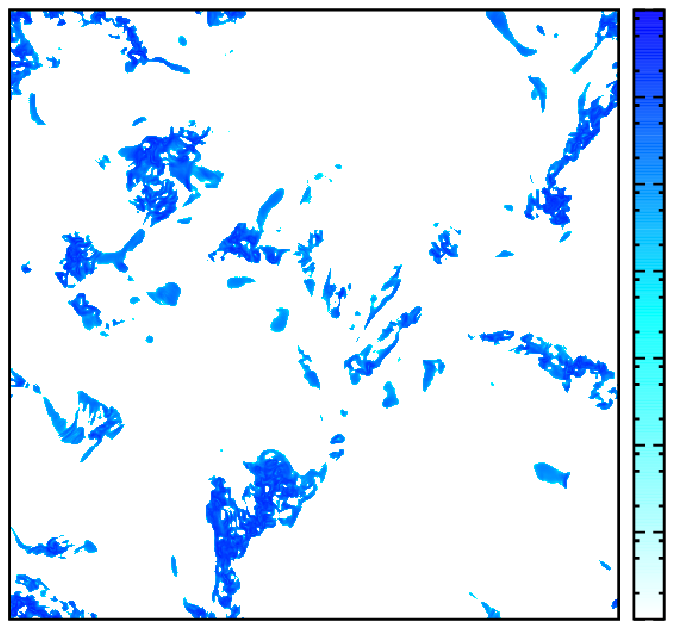} &
\input{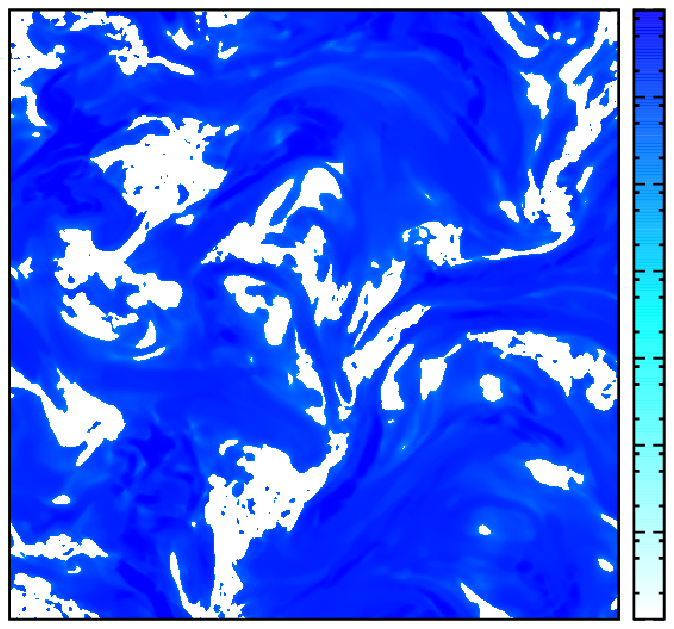} \\
\input{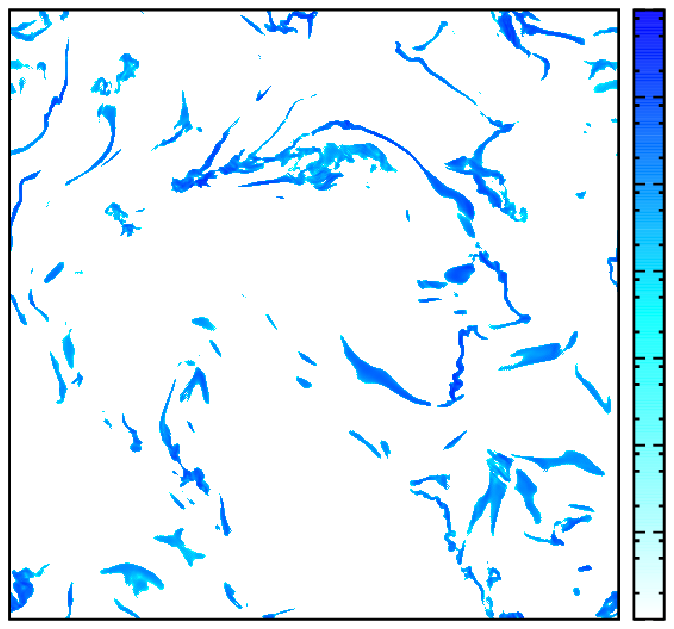} &
\input{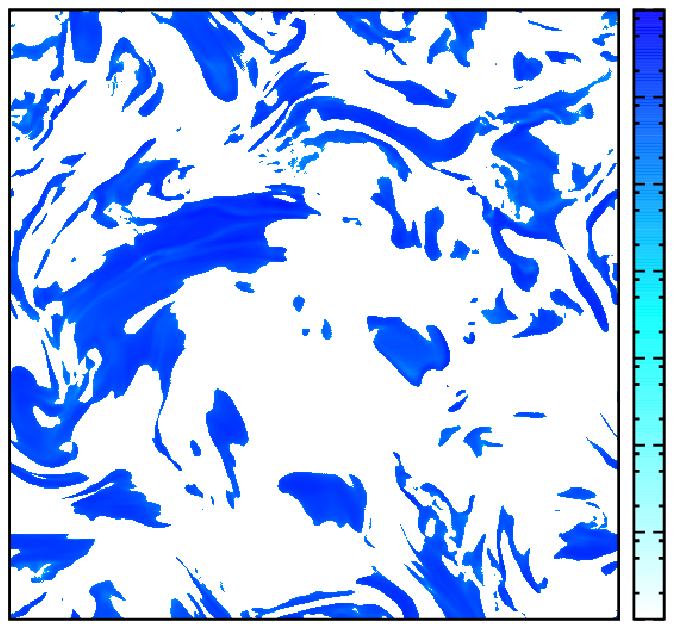} \\
\input{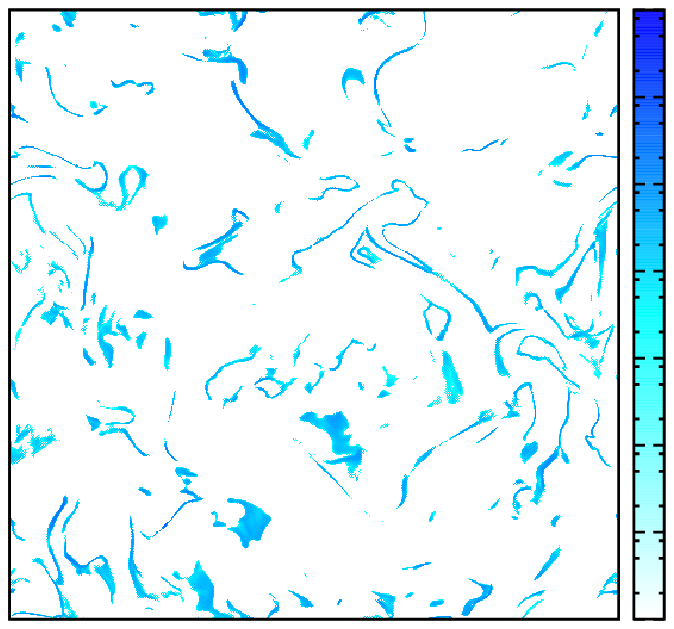} &
\input{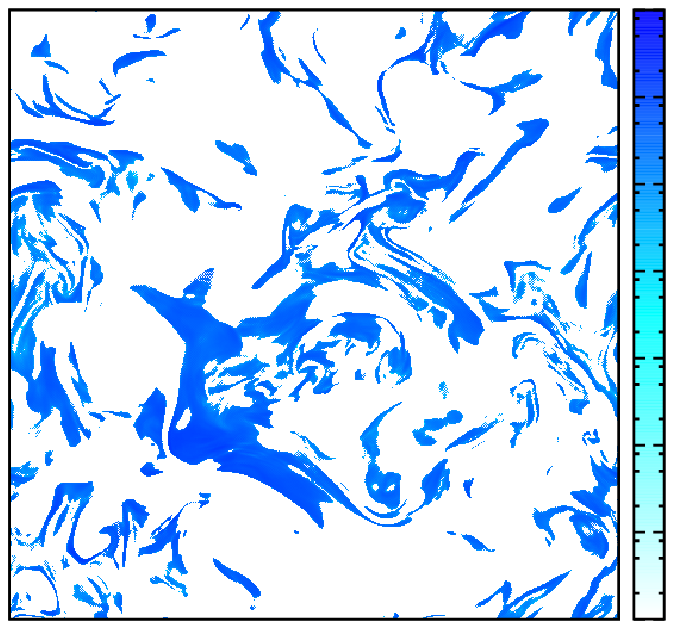}
\end{tabular}
\caption{Central slice in the XY plane of the domain showing distributions of the maximum growth rate $\gamma_{max}$ (normalized by the initial ion gyrofrequency $\Omega_{i0}$) of the firehose (left column) and mirror (right column) instabilities for models A (with $\beta_0=200$ and different values of the anisotropy relaxation rate $\nu_S$).  The expressions for the maximum growth rates are given by Equations~\eqref{eqn:max_growth_rates}, 
with a maximum value given by 
 $\gamma_{max}/\Omega_i=1$. Data are taken at the the final time $t_{f}$ for each model, indicated in Table~\ref{tab:models}. }
\label{fig:fhmi}
\end{figure*}

As an illustration of the spatial distribution of the unstable gas, Figure~\ref{fig:fhmi} depicts  maps of the maximum growth rate of  both the firehose (left column) and the mirror (right column) instabilities given by Equations~\eqref{eqn:max_growth_rates} for  the models with moderate initial magnetic field ($\beta_0=200$) and different anisotropy relaxation rates $\nu_S$.\footnote{We note that because Equations~\eqref{eqn:max_growth_rates} have a validity limit as described in \S 2.1, we have corrected the growth rates to $\gamma_{max}/\Omega_i = 1$  when  outside of the validity range. This limit is well justified by fully solutions of the dispersion relation obtained from the linearization of the Vlasov-Maxwell equation by Gary (1993; see Chapter 7). }
These maximum growth rates are normalized by the initial ion gyrofrequency $\Omega_{i0}$ and occur for modes with wavelengths of the order of the ion Larmor radius. 
First thing to note is that the mirror unstable regions have a larger volume filling factor than the firehose unstable regions for all the models in Figure~\ref{fig:fhmi}.
 This is because the regions where the magnetic field is amplified have a large perpendicular pressure and this happens on most of the turbulent volume.  Regions with an excess of parallel pressure arise when the magnetic intensity decays, like in regions with magnetic field reversals. The correspondence of the low intensity magnetic field with firehose unstable regions can be checked directly in model A2 by comparing the  maps of  Figures~\ref{fig:fhmi} and~\ref{fig:maps_dens}. The firehose unstable regions in model A2 in Figure~\ref{fig:fhmi} are  small  and fragmented; while  in models A3 and A4,  they are elongated (at lengths of the turbulent injection scale) and very thin (with  thickness  of the order of the dissipative scales) and are regions with magnetic field reversals and reconnection.

Also, from   Figure~\ref{fig:fhmi} we see that most of the volume of A2 model is mirror unstable; for models A3 and A4, the mirror unstable regions are elongated but with much larger thickness than in the firehose unstable regions. 
We must remember that the criterium  for the mirror unstable regions in Figure~\ref{fig:fhmi} is the kinetic one (Eq. 9) rather than the CGL-MHD criterium (Eq. 8) (see also Figure~\ref{fig:beta_vs_a}).

The spatial dimensions of the unstable regions in Figure~\ref{fig:fhmi} also reveal the maximum wavelength
 of the unstable modes which should develop inside the turbulent domain. 
 In the models with finite anisotropy relaxation rate $\nu_S$, the larger the value of  $\nu_S$ the smaller the wavelength of the unstable modes. For realistic values of $\nu_S$ of the order of $\gamma_{max}$ (the maximum frequency of the instabilities), there would have only unstable modes with wavelengths bellow the spatial dimensions we can solve.

\subsection{Magnetic versus thermal stresses}

The gyrotropic tensor gives the gas a larger (smaller) strength to resist against bending or stretching of the field lines if $A>1$ ($A<1$). This higher or smaller strength comes from the parallel anisotropic force
\begin{equation}
f_A = (p_{\parallel} - p_{\perp}) \nabla_{\parallel} \ln{B} ,
\end{equation}
where $\nabla_{\parallel} \equiv (\mathbf{B}/B) \cdot \nabla$. The relative strength between this anisotropic force and the usual Lorentz curvature force can be estimated from $\alpha \equiv (p_{\parallel} - p_{\perp}) / (B^{2} / 4\pi)$.

As a measure of the dynamical importance of the anisotropy, we calculated the average value of 
$| \alpha |$ for all the models of Table~\ref{tab:models} and the values are listed in Tables~\ref{tab:statistics_A}, ~\ref{tab:statistics_B}, and~\ref{tab:statistics_C} for models A, B and C, respectively.

First let us consider the models  with initial moderate magnetic field ($\beta_0=200$). For models A1, A5, and A6, with instantaneous relaxation of the anisotropy to the marginally stable state, the anisotropic force is non dominant: $\langle |\alpha| \rangle \approx 0.4$. For  model A2, with no relaxation on the anisotropy (standard CGL-model), on the other hand, the anisotropic force is dominant, with $\langle |\alpha| \rangle \approx 5$. For the  models A3 and A4, with finite isotropization rate, the anisotropic force is comparable to the curvature force, being smaller for the higher isotropization rate: $\langle |\alpha| \rangle \approx 3$ for model A3 ($\nu_S=10^{2}$) and $\langle |\alpha| \rangle \approx 1.5$ for model A4 ($\nu_S=10^{3}$).

For  model  with strong magnetic field ($\beta_0=0.2$) B1, the anisotropic force is negligible compared to the Lorentz curvature force: $\langle |\alpha| \rangle \approx 0.04$.

\subsection{PDF of Density}

\begin{figure}
\centering
\begin{tabular}{c c}
\input{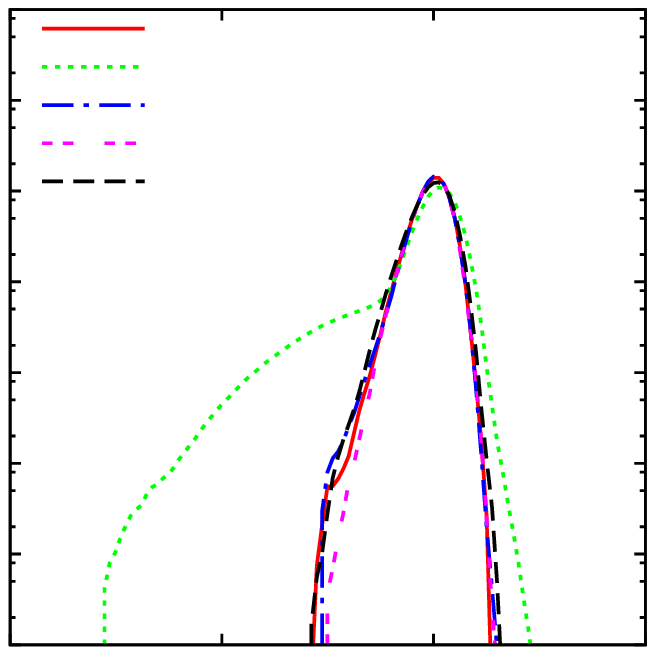} \\
\input{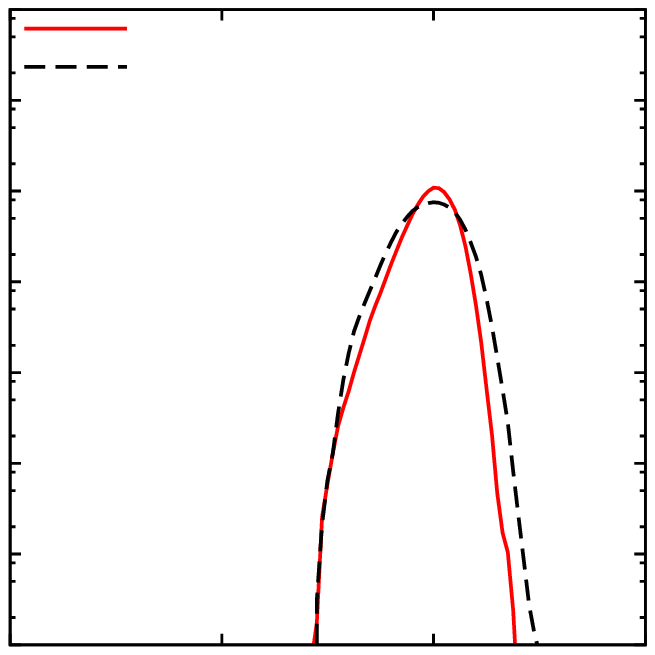}
\end{tabular}
\caption{Normalized histogram of $\log{\rho}$. Top: models starting with $\beta_0=200$. Bottom: models starting with $\beta_0=0.2$. The histograms were calculated using one snapshot every $\Delta t = 1$, from $t=2$ until the final time $t_{f}$ indicated in Table~\ref{tab:models}.}
\label{fig:pdf_ldn}
\end{figure}

Figure~\ref{fig:pdf_ldn} shows the normalized histograms of $\log{\rho}$ for  models  A and B of Table~\ref{tab:models} having different rates of anisotropy relaxation $\nu_S$. 
The upper panel shows models with initial moderate magnetic field intensity ($\beta_0=200$) and the lower panel the model  with initial strong magnetic field intensity ($\beta_0=0.2$). The corresponding collisional MHD models are also shown for comparison. 

Examining the high $\beta$  models in the top diagram, we note  that all the models with anisotropy relaxation have  similar  distribution to the collisional model.
Model A2, for which the anisotropy relaxation is null, has a much broader distribution, specially in the low density domain.
This difference is due to the presence of strong mirror forces in the A2 model which expels  the gas to outside of high magnetic field intensity regions, causing the formation of low density zones. 
Consistently, we can check this effect in the bi-histograms of  density versus magnetic field intensity in Figure~\ref{fig:pdf_ldn_lb} of Appendix for model A2 (the bi-histogram for the model Amhd is also shown for comparison). The lowest density points are correlated with high intensity magnetic fields for the model A2. 

The bottom  panel of Figure~\ref{fig:pdf_ldn} indicates that the low $\beta$, strong magnetic field model B1 has density distribution only slightly narrower than the collisional MHD model Bmhd, specially at the high density region. 
The slight difference with respect to the collisional model is possibly due to: (i) the sound speed parallel to the field lines is higher in the collisionless models:  $c_{\parallel s} = \sqrt{3 p_{\parallel}/\rho}$ for the collisionless model, while for the collisional model $c_s = \sqrt{5 p/3 \rho}$; and (ii) in the direction perpendicular to the magnetic field, the fast modes have characteristic speeds higher in the collisionless model: $c_f = \sqrt{B^{2}/4\pi\rho + 2p_{\perp}/\rho}$, while for the MHD model $c_f =  \sqrt{B^{2}/4\pi\rho + 5p/3\rho}$. These larger speeds in the anisotropic model imply a larger resistance to compression and therefore, smaller density enhancements (at least for our transonic models).

\subsection{The turbulence power spectra}

Power spectrum is an important characteristic of turbulence. For MHD turbulence a substantial progress
has been achieved recently as the Goldreich-Sridhar model has become acceptable. Recent numerical
work 
has tried to  resolve the controversies and 
confirmed the Kolmogorov $-5/3$ spectrum
of Alfv\'enic turbulence predicted in the model
(e.g. \citealt{beresnyak_lazarian_2009, beresnyak_lazarian_2010, beresnyak_2011, beresnyak_2012b}). 
This spectrum corresponds to the Alfv\'enic mode of
the compressible MHD turbulence (\citealt{cho_lazarian_2002, cho_lazarian_2003, kowal_lazarian_2010}, Beresnyak \& Lazarian 2013). 

Our goal here is to determine the power spectrum of the turbulence
in collisionless plasma in the presence of the feedback of plasma instabilities on scattering.

\begin{figure*}
\centering
\begin{tabular}{c c}
\input{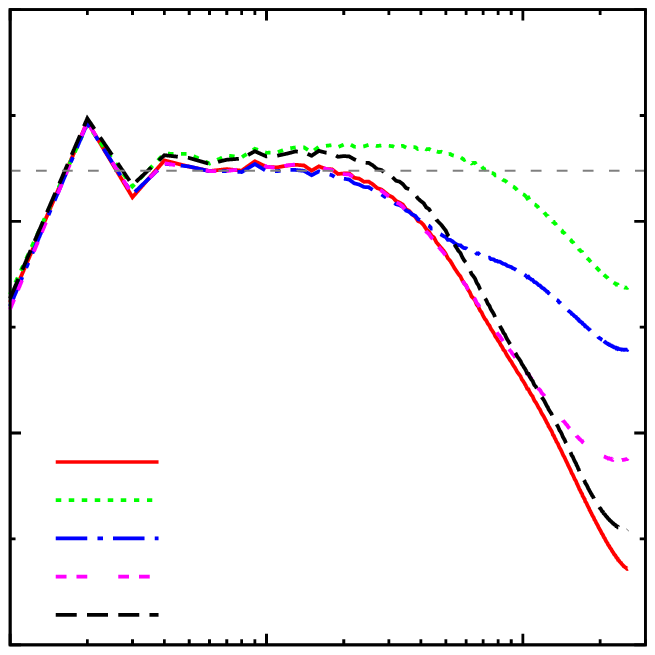} &
\input{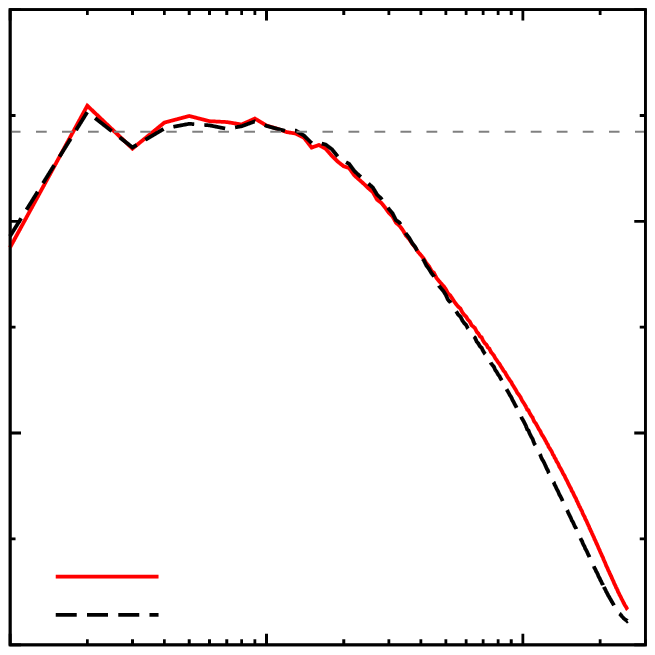} \\
\input{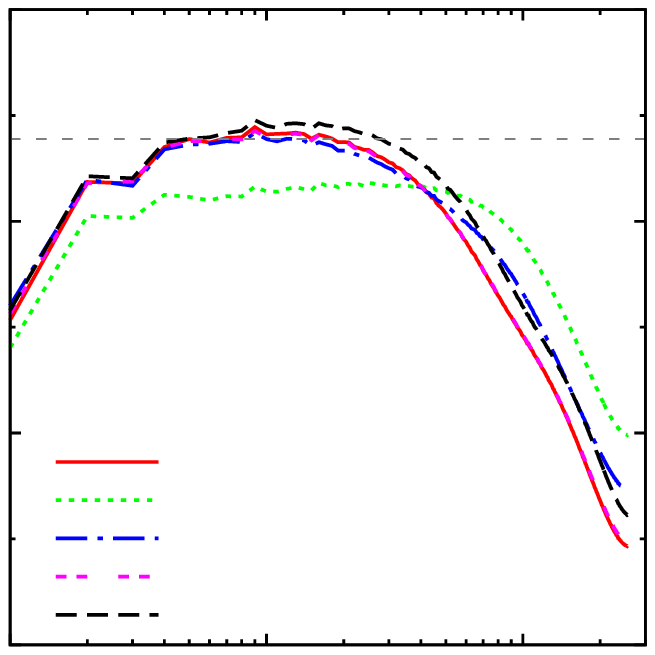} &
\input{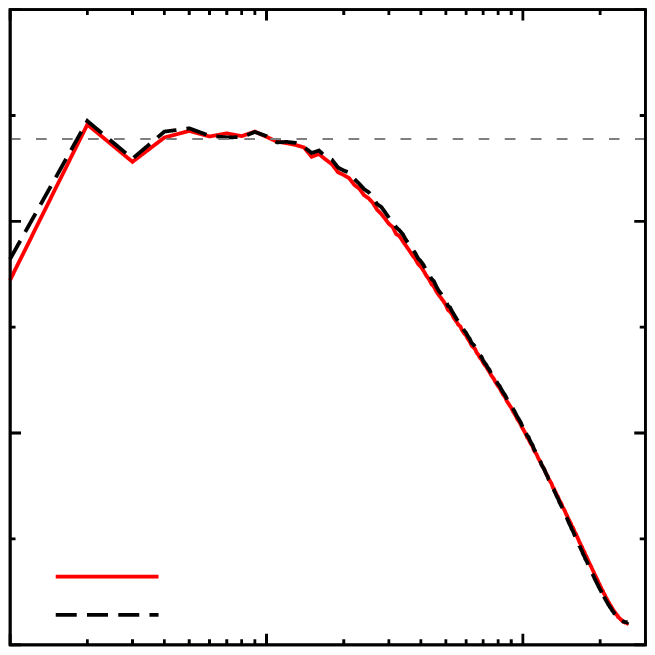} \\
\input{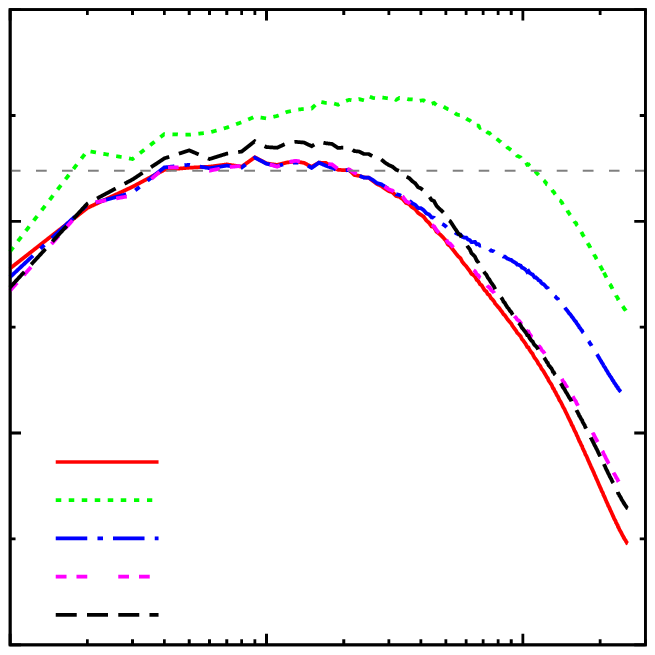} &
\input{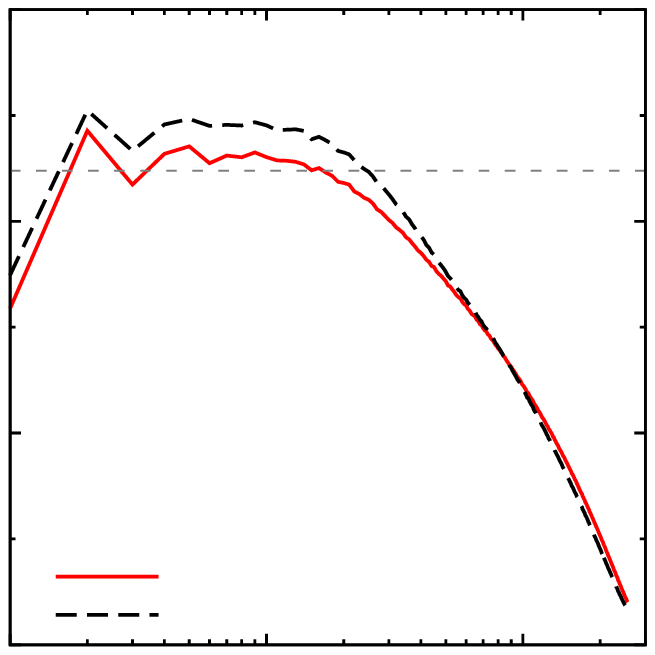}
\end{tabular}
\caption{Power spectra of the velocity $P_u(k)$ (top row), magnetic field $P_B(k)$ (middle row), and density $P_{\rho}(k)$ (bottom row), multiplied by $k^{5/3}$. Left column: models A,  with initial $\beta_0 = 200$. Right column: models B, with $\beta_0=0.2$.  Each power spectrum was averaged in time considering snapshots every $\Delta t = 1$, from $t=2$ to  the final time step  $t_{f}$ indicated in Table~\ref{tab:models}.}
\label{fig:ps_turb}
\end{figure*}

Figure~\ref{fig:ps_turb} compares, for different models of Table~\ref{tab:models}, the power spectra of the velocity (top row), magnetic field (middle row) and density (bottom row). The models starting with moderate magnetic field and $\beta_0=200$ (A1, A2, A3, A4, Amhd), for which the turbulence is super-Alfv\'enic, are in the left column, and the models starting with strong magnetic field and $\beta_0=0.2$ (B1, Bmhd), for which the turbulence is sub-Alfv\'enic are in the right column. Each power spectrum is multiplied by the factor $k^{5/3}$.

The velocity power spectrum $P_u(k)$ for the super-Alfv\'enic high beta collisional model Amhd (in the left top panel of Figure~\ref{fig:ps_turb}) is consistent with the Kolmogorov slope in the approximate interval $4<k<20$ and decays quickly for $k>30$. The power spectrum of the collisionless models A1, A3, and A4 are similar, but  show slightly less power in the interval $4<k<30$. In fact, in Table~\ref{tab:statistics_A} in the Appendix, we find that the average values of $u^{2}$ for these models are smaller than the model Amhd. Model A3 evidences more power at the smallest scales, already at the dissipation range. This is due to the acceleration of gas produced by the firehose instability (see fig.~\ref{fig:beta_vs_a}). Model A2, which has null anisotropy relaxation rate (standard CGL model), has a flatter velocity power spectrum than the collisional MHD model Amhd, and much more power at the smallest scales. This excess of power comes from the firehose and mirror instabilities and is consistent with the trend reported  in the previous sections and also in  \cite{kowal_etal_2011a}.

The sub-Alf\'enic velocity power spectrum  $P_u(k)$ of the collisional MHD model Bmhd (top right panel in Figure~\ref{fig:ps_turb}) has a narrower interval of wavenumbers consistent  with the Kolmogorov slope. The power spectra   $P_u(k)$ of the collisionless model B1 is almost identical which is in agreement with the small dynamical importance of the anisotropy  forces compared to the magnetic forces (see \S 4.2).

The power spectrum related to the compressible component of the velocity field $P_C(k)$ is shown in Figure~\ref{fig:ps_velcvelo} in the Appendix, where it is divided at each wavenumber by the total power of the velocity field. For the high beta models, the ratio $P_C(k)/P_u(k)$ for the collisionless models is similar to that of the collisional MHD model Amhd for almost every wavenumber $k$ and is $\approx 0.15$. 
For the low beta model, however, the collisionless model has a ratio   $P_C(k)/P_u(k)$ slightly higher than that of the collisional MHD model Bmhd for wavenumbers above $k\approx10$.
The fractional power in the compressible modes in the interval $2<k<10$  is smaller compared to the super-Alfv\'enic (high $\beta$) models, but at larger wavenumbers it becomes higher.

\begin{figure}
\centering
\input{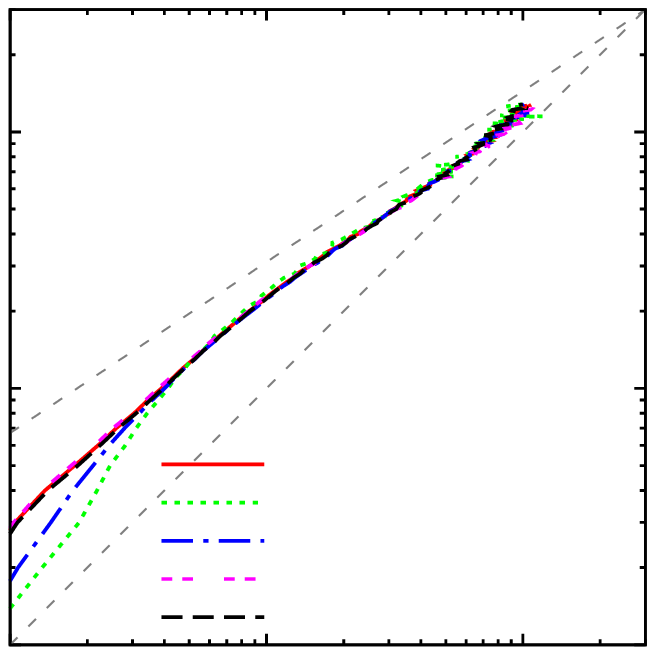} \\
\input{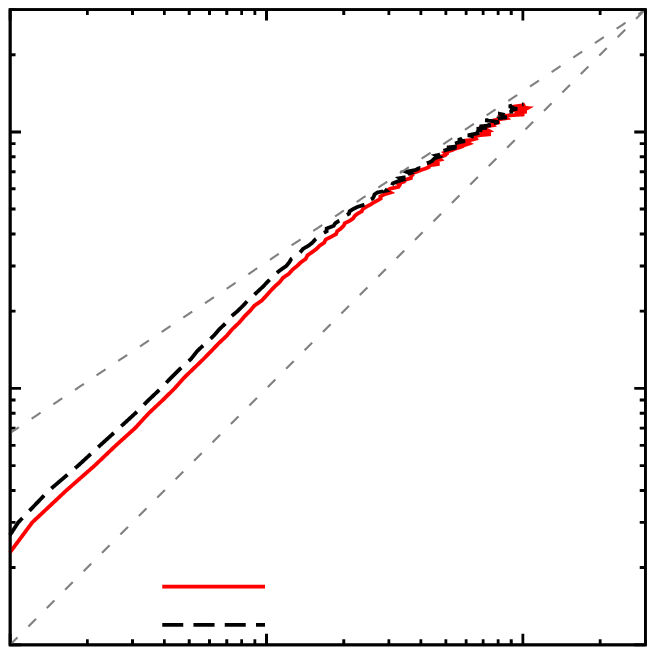}
\caption{$l_{\perp}$ vs $l_{\parallel}$ obtained from the structure function of the velocity field (Eq.~\ref{eqn:sf2}). The axes are scaled in cell units.}
\label{fig:sf2_velo}
\end{figure}

The anisotropy in the structure function of the velocity is shown in Figure~\ref{fig:sf2_velo}. The structure function of the velocity $S^{u}_{2}$ is defined by
\begin{equation}
S^{u}_{2}(l_{\parallel}, l_{\perp}) \equiv \langle | \mathbf{u(r+l)} - \mathbf{u(r)} |^{2} \rangle,
\label{eqn:sf2}
\end{equation}
where the displacement vector $\mathbf{l}$ has the parallel and perpendicular components (relative to the local mean magnetic field) $l_{\parallel}$ and $l_{\perp}$, respectively. The local mean magnetic field is defined by $(\mathbf{B(r+l) + B(r)})/2$ (like in \citealt{zrake_macfadyen_2012}). The GS95 theory predicts an anisotropy scale dependence of the velocity structures (eddies) of the form $l_{\parallel} \propto l_{\perp}^{2/3}$. 
The axis in Figure~\ref{fig:sf2_velo} are in cell units. The collisional MHD model Amhd is consistent with the GS95 scaling for the interval $10 \Delta<l_{\perp}<40 \Delta$, where $\Delta$ is one cell unit in the computational grid. For the sub-Alfv\'enic model Bmhd, however, this scaling is less clear, although the anisotropy is clearly seen.

The collisionless models A1, A3, and A4 in Figure~\ref{fig:sf2_velo} (which have pressure anisotropy relaxation) evidence anisotropy in the velocity structures which is identical to that of the collisional MHD model Amhd. Model A2 (with no pressure anisotropy constraint), on the other hand, has more isotropized structures at  small values of $l$. This effect is due to the action of the instabilities and is also observed in the high beta  models in \cite{kowal_etal_2011a} for both the firehose and mirror instabilitiy regimes.

The magnetic field power spectra $P_B(k)$ of the collisional MHD models Amhd and Bmhd (middle row of Figure~\ref{fig:ps_turb}) show a power law consistent with the Kolmogorov slope at the same intervals  of the velocity power spectra. As in the velocity power spectrum, in the high beta, super-Alfv\'enic cases, the collisionless models A1, A3, and A4 have similar $P_B(k)$ to the collisional model Amhd (although  with slightly less power). 
Model A2 has a $P_B(k)$  much flatter than that of Amhd and has less power (by a factor of two) at the inertial range interval. In the smallest scales ($k>50$), however, its power is above that of the Amhd model. As in the velocity power spectrum, these  small-scale structures are due to the instabilities which are present in this model.

For the sub-Alfv\'enic, low beta models (B), the magnetic field power spectrum $P_B(k)$ of the collisionless model is again similar to the collisional MHD model Bmhd.

Figure~\ref{fig:ps_magnvelo} of the Appendix compares $P_B(k)$ and $P_u(k)$ for our models. For the super-Alfv\'enic, high beta models (A) which are in steady state, the magnetic field power spectrum is in super equipartition with the velocity power spectrum for $k>3$ for all models, but the A2 model which has $P_B(k)<P_u(k)$ for all wavenumbers. Models A3 and A4 show $P_B(k)/P_u(k)$ decreasing values for larger wavenumbers, being this effect more pronounced in  model A3 which has smaller anisotropy relaxation rate. The sub-Alfv\'enic, low beta collisionless model B1 has the ratio $P_B(k)/P_u(k)$ slightly smaller than unity for all wavenumbers and slightly smaller than the collisional Bmhd model at large $k$ values.

The anisotropy in the structure function for the magnetic field shows similar trend to the velocity field in all models and is not presented here.
Likewise, the density power spectra  $P_{\rho}(k)$ for the super-Alfv\'enic, high beta models (bottom row in Figure~\ref{fig:ps_turb}) reveal the same trend of  the velocity power spectra.
For the sub-Alfv\'enic model, however, the smaller power in the larger scales compared to the collisional MHD model Bmhd is clearly evident, specially in the inertial range. This is consistent with the discussion following the presentation of the density distribution (\S 4.3), which evidenced that the collisionless models resist more to compression than the collisional model (see also Figure~\ref{fig:ps_velcvelo}).

\subsection{Turbulent amplification of seed magnetic fields}

\begin{figure*}
\centering
\begin{tabular}{c c}
\input{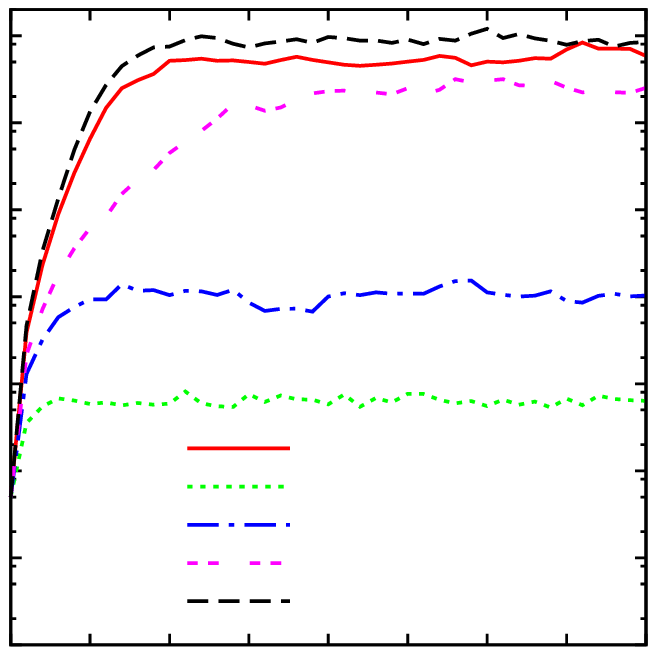} &
\input{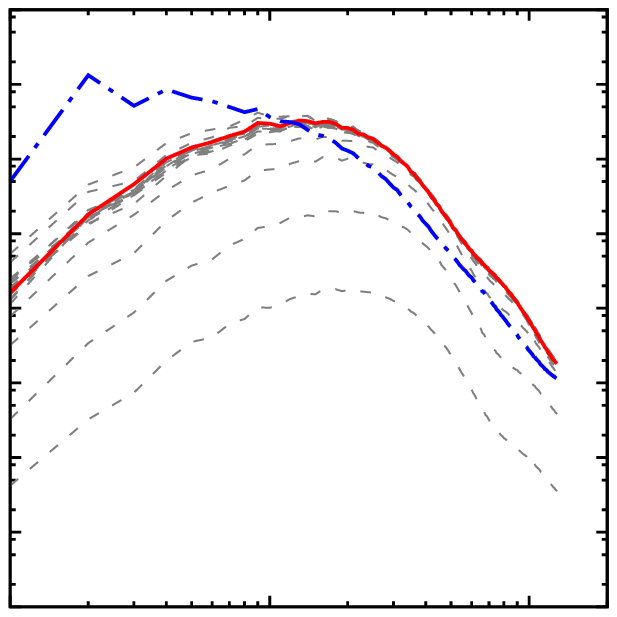} \\
\input{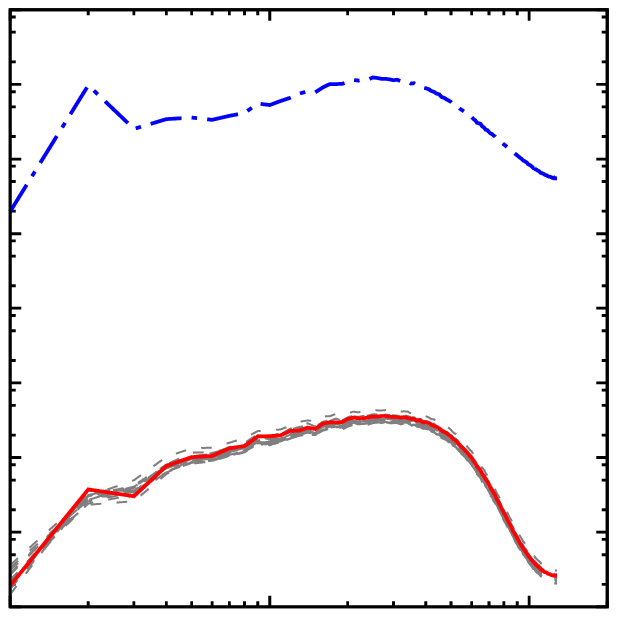} &
\input{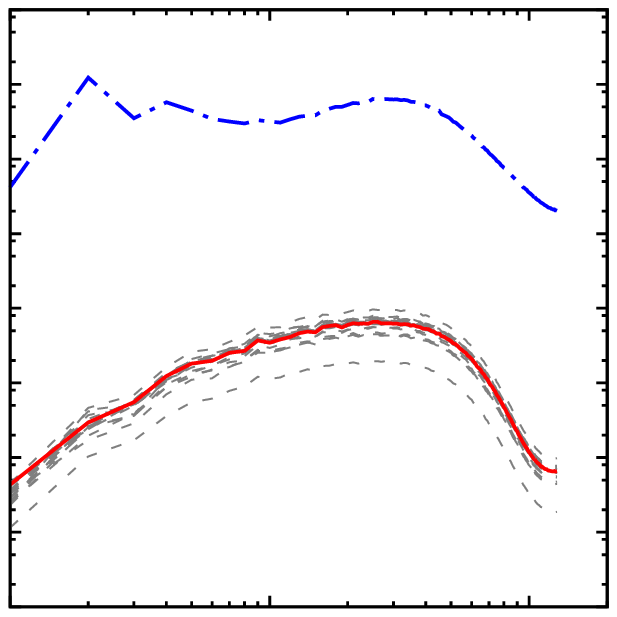} \\
\input{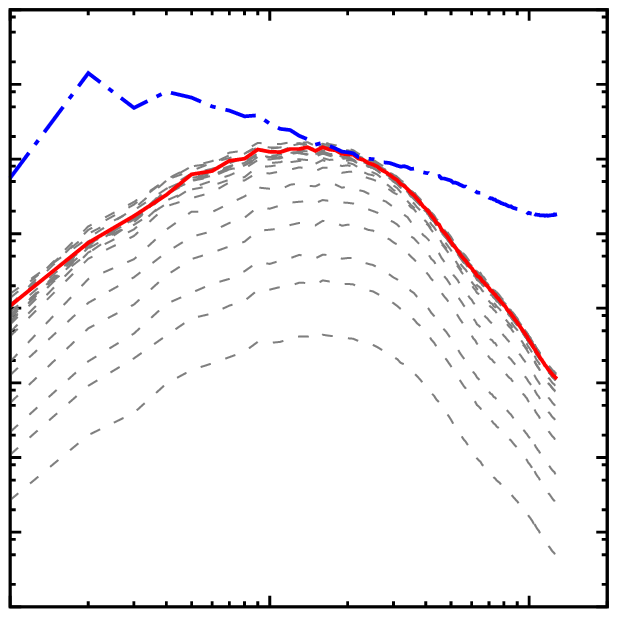} &
\input{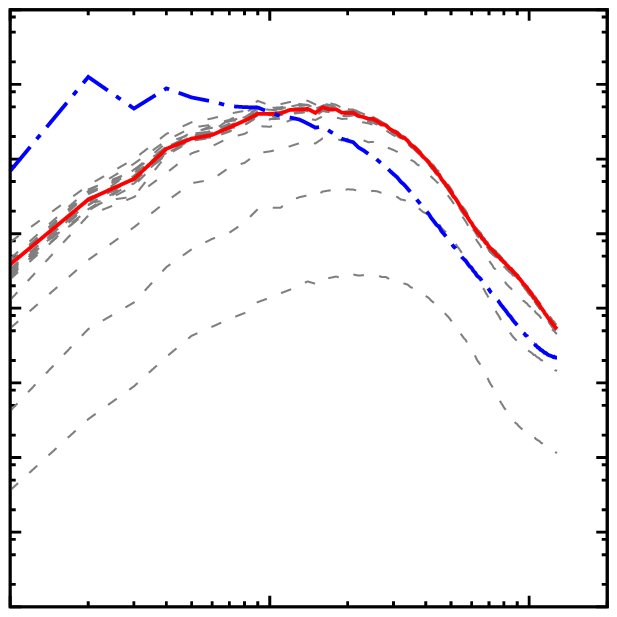}
\end{tabular}
\caption{Top left: time evolution of the magnetic energy $E_M=B^{2}/2$ for the models starting with a weak (seed) magnetic field, models C1, C2, C3, C4, and Cmhd, from Table~\ref{tab:models}. The other panels show the magnetic field power spectrum multiplied by $k^{5/3}$ for each model, from $t=2$ at every $\Delta t =2$ (dashed lines) until the final time indicated in Table~\ref{tab:models} (solid lines). The  velocity field power spectrum multiplied  by $k^{5/3}$ at the final time is also depicted for comparison (dash-dotted line).}
\label{fig:evol_dyn}
\end{figure*}

Top left panel in Figure~\ref{fig:evol_dyn} shows the magnetic energy evolution of the models having initially very weak magnetic  (seed) field, models C1, C2, C3, C4, and Cmhd of Table~\ref{tab:models}. The kinetic energy of the models is not shown, but their values are approximately constant in time (after $t\approx1$) and their average values (taken during the last $\Delta t = 10$ for each model)  $\langle E_K \rangle$ are shown in Table~\ref{tab:statistics_C}. 
The other panels in Figure~\ref{fig:evol_dyn} show the power spectrum of the magnetic field, from $t=2$ until the final time, for every $\Delta t = 2$ (dashed lines). The final magnetic field power spectrum is the continuous line. Also for comparison, it is plotted the final velocity power spectrum (dash-dotted line).

 The collisional MHD  model Cmhd shows an initial exponential growth of the magnetic energy until $t \approx 10$. In this interval, the average magnetic energy grows from $E_M=5 \times 10^{-7}$ to $E_M \sim 10^{-2}$. 
At the final times, the magnetic energy achieves the value $E_M \approx 9.0 \times 10^{-2}$ which is approximately four times smaller than the average kinetic energy $E_K \approx 0.38$ (see Table~\ref{tab:statistics_C}). Bottom right panel of Figure~\ref{fig:evol_dyn} shows that the final magnetic field power spectrum is peaked at $k\approx 20$ above  which it is in super-equipartition with the velocity power spectrum.

The collisionless model C1 with instantaneous relaxation of the pressure anisotropy ($\nu_S=\infty$), 
has a turbulent amplification of  the magnetic energy very similar to that of the collisional MHD model Cmhd (top left panel in Figure~\ref{fig:evol_dyn}). The initial exponential growth rates are indistinguishable between the two models, but  the final value of saturation of the magnetic energy is slightly smaller: $E_M \approx 6.2 \times 10^{-2}$ (see Table~\ref{tab:statistics_C}). During the initial exponential growth of the magnetic energy, when the plasma still  has high values of $\beta$, the pressure anisotropy relaxation due to the kinetic instabilities keeps the plasma mostly isotropic, explaining the similar behaviour to the collisional MHD model. When $\beta$  starts to decrease, the anisotropy $A$ can increase (or decrease) spanning  a range of $A$ values in the stable zone 
(as in Figure 2 for model A1). Then, the anisotropic forces can start to have  dynamical importance. At the final times, the value of $ \langle |p_{\parallel}-p_{\perp}| / (B^{2}/4\pi)\rangle$, which measures the dynamical importance of the anisotropic forces compared to the Lorentz curvature force (see \S 4.2) is $\approx0.5$ (Table~\ref{tab:statistics_C}). The magnetic field power spectrum has an identical shape to the model Cmhd, specially in the final time step.

The turbulent dynamo is also tested for a model with a finite anisotropy relaxation rate,  model C4, which has $\nu_S=10^{2}$. The growth rate of the magnetic energy in the exponential phase is smaller compared to models Cmhd and C1 (top left panel in Figure~\ref{fig:evol_dyn}). In this case, the anisotropy $A>1$ develops moderately during the magnetic energy amplification and gives the mirror forces some dynamical importance to change the usual collisional MHD dynamics. The value of the magnetic energy at the final time of the simulation is approximately one third of the value for the collisional MHD model. The final magnetic power spectrum has a shape similar to the collisional MHD model Cmhd, but below the equipartition with the velocity field power spectrum which has more power at the smallest scales due to the presence of the instabilities.

Model C2, a standard CGL model with no constraints on the growth of  pressure anisotropy ($\nu_S=0$) shows no evidence of a turbulent dynamo amplification of its magnetic  energy which saturates at very low values already at $t\approx 5$  (Figure~\ref{fig:evol_dyn}), when $E_M \approx 6.2 \times 10^{-6}$, while the kinetic energy is  $E_K \approx 0.32$ (see Table~\ref{tab:statistics_C}). The reason is that the anisotropy $A$ increases at the same time that the magnetic field is increased ($A \propto B^{3}/\rho^{2}$ in the CGL closure), giving rise to strong mirror forces along the field lines which increase their resistance against  bending or stretching. For this model, $ \langle |p_{\parallel}-p_{\perp}| / (B^{2}/4\pi)\rangle \sim 10^{5}$, that is, the anisotropic forces dominate over the Lorentz force. 
The magnetic field power spectrum (middle left panel in Figure~\ref{fig:evol_dyn}) is similar in shape (but not in intensity) to the Cmhd model, being peaked at $k \approx 40$.

The saturated value of the magnetic energy for  models without anisotropy relaxation is, nevertheless, sensitive to the initial plasma $\beta$.  Model C3 is similar to model C2, but starts with a lower sound speed ($V_{S0} = 0.3$) which makes $\beta$ ten times smaller (see Table 1). Turbulence is supersonic in this case, rather than transonic. The magnetic energy evolution is similar to that of the model C2, but the magnetic energy saturates with a value about two orders of magnitude larger, although  the anisotropic forces are still dominant, with   $ \langle |p_{\parallel}-p_{\perp}| / (B^{2}/4\pi)\rangle \sim  10^{4}$ (see Table~\ref{tab:statistics_C}).

\section{Discussion}

 In order to study the collisionless effects in the turbulence statistics and also in the magnetic field amplification in the ICM plasma, we performed 3D numerical simulations of forced turbulence (continuously injected) in a periodic box, employing a single-fluid collisionless MHD-type model described by Equations \eqref{eqn:collisionless_mhd}. These are similar to the single fluid CGL-MHD equations (\citealt{chew_etal_1956}), but modified by source terms which, in particular, account for the relaxation of the pressure anisotropy that arises from the collisionless MHD condition. We adopt a model of the relaxation that is motivated
by both theoretical considerations and measurements of solar wind and laboratory plasmas. The gist of
our approach is based on the fact that plasma instabilities induce scattering of particles and this decreases the mean free
paths of the particles and isotropize their distribution. As the exact measures of scattering are difficult to obtain from the first principles, we adopt an empirical approach making sure that the distribution
is consistent with the measurements in solar wind and laboratory plasmas.  

To better explore the parameter space we have also considered different values of the pressure anisotropy relaxation rate $\nu_S$. Models with $\nu_S=\infty$ (i.e. with instantaneous relaxation rate)  represent conditions for which the relaxation time $\sim \nu_S^{-1}$ is much shorter than the minimum time step $\delta t_{min}$ that our numerical simulations  are able to solve ($\delta t_{min} \sim 10^{-6}$). 
Previous studies (\citealt{gary_etal_1997, gary_etal_1998, gary_etal_2000}) suggest that the rate $\nu_S$ should  be of the order of a few percent of $\Omega_p$, the proton gyrofrequency (see discussion below). If we consider typical physical conditions for the ICM, in order to convert the code units into physical units (see \S 3.2), we may take  $l_* = 100$ kpc, $v_* = 10^{8}$ cm/s, and $\rho_* = 10^{-27}$ g/cm$^{3}$ as characteristic values for the length scale, dynamical velocity and density of the ICM, respectively. This implies a characteristic time scale $t_{*} \sim 10^{15}$ s, while for  models A in Table~\ref{tab:models}, the proton Larmor period is $\tau_{cp} \sim 10^3$ s. Using  $\nu_S \sim 10^{-3} \tau_{cp}^{-1}$, we find $\nu_S^{-1} \sim 10^{-9} t_*$. Therefore, the models of Table 1  for which we assumed  $\nu_S = \infty$ are very good approximations to the description of the direct effect of plasma instabilities at the large scale turbulent
motions within ICM.
For comparison, we have also run models with no anisotropy relaxation, or $\nu_S=0$, which thus behave like  standard CGL-models.

Employing  the same thermal sound  speed and turbulent velocity (with a turbulent sonic Mach number $M_S \approx 1$), we simulated models having different initial magnetic field intensities, or different plasma $\beta$ parameters (see Table~\ref{tab:models})\footnote{We remark that model C3 was  the only one with a smaller thermal speed, implying a supersonic turbulent regime}. 
In particular, models with  high beta  ($\beta_0=200$) are representative of typical ICM conditions.

Let us  first summarize the results for the super-Alfv\'enic, high $\beta$  models (models A of Table~\ref{tab:models}). We have found that for  model A1 ($\nu_S=\infty$) the turbulent properties (density, velocity and magnetic field power spectra, and the anisotropy in the structure functions of these fields) are qualitatively and quantitatively similar to the collisional MHD model Amhd. The dynamical importance of the anisotropic force (Eq. 16, in comparison to the Lorentz force) is small in this case. The models with finite values of $\nu_S$, A3 ($\nu_S = 10^{2}$) and  A4  ($\nu_S = 10^{3}$) allow the plasma to develop kinetic instabilities in small zones of the domain, which are larger  for model A3. Comparing with model A1, the anisotropic forces of these models are stronger, but the turbulence statistics does not change significantly at large scales. The main differences are seen at small scales, where the instabilities growth rates are larger. 

Model A2  ($\nu_S=0$) showed a clear deviation from the results of the collisional model Amhd. In this case, practically all the domain is unstable to the kinetic (mostly mirror) instabilities, causing accumulation of power in density and velocity fluctuations at the smallest solved scales. In this case, these fluctuations are damped only by  numerical dissipation. At the same time, the power in the magnetic fluctuations in large and moderate scales is reduced. The field lines resist more to bending  (in most of the volume for which $A>1$) due to the dominance of the mirror forces.

The sub-Alfv\'enic, low $\beta$ model B1 (of Table~\ref{tab:models}) is insensitive to the anisotropy relaxation rate $\nu_S$ as the turbulence creates anisotropies, but no kinetic instabilities in this case (at least not in the transonic turbulent regime investigated here). We have seen that the anisotropic forces are negligible compared to the Lorentz force in this case and the turbulence statistics is well approximated by the collisional MHD  description, as in model Bmhd.

Our models starting with seed magnetic fields (models C of Table~\ref{tab:models}) evidence the dramatic effect of the rate $\nu_S$ on the turbulent amplification of the magnetic field. Model C1 ($\nu_S=\infty$) produces a magnetic amplification rate similar to the collisional MHD case Cmhd. During the exponential phase of magnetic energy amplification, the plasma $\beta$ value is high and the  relaxation of the anisotropy keeps the pressure isotropic, so that  the dynamics is similar to that of the collisional MHD. When the magnetic field has enough energy and  $\beta$ decays to values close to unity, then the anisotropic forces begin to compete with the Lorentz force. Even though, the turbulence statistics at the final state of saturation of the  magnetic energy (or close to it) is similar to the collisional MHD model, Cmhd. Compared to model Cmhd, the magnetic energy of model  C1 achieves a saturation value around $30\%$ smaller only.

Models C2 and C3, which have no constraint on anisotropy growth  ($\nu_S=0$) fail to amplify the magnetic energy. The mirror forces are large and make the plasma to resist to the stretching of the field lines by the turbulent motions. The magnetic energy at the final steady state is several orders of magnitude smaller than the kinetic energy. This is consistent with earlier results presented in~\citet{santos-lima_etal_2011} and~\citet{dgdp_etal_2013}, and also with the  findings of~\citet{santos-lima_etal_2009}, where the failure of the turbulent dynamo using a double-isothermal closure for $p_{\perp} > p_{\parallel}$ was  reported.
Model C4  with finite anisotropy relaxation rate ($\nu_S=10^{2}$), on the other hand, has a growth rate of the magnetic energy only a little smaller than the  Cmhd and C1 models during the exponential amplification and the magnetic energy at the  final saturation  state is about one third of that of the Cmhd.

In summary, all our collisionless models with $\nu_S=\infty$ have exhibited no significant deviations from the collisional models in the turbulent power spectra of density, magnetic field and velocity, or in the anisotropy of the structure function of these fields. 
Models with finite large enough values of $\nu_S$ also reveal statistical behaviour that approaches the collisional solutions.

\subsection{Further considerations about the instabilities and the anisotropy relaxation models}

The anisotropy in pressure created by the turbulent motions gives rise to new forces in the collisionless MHD description (see the momentum conservation equation in Eq.~\ref{eqn:collisionless_mhd}). These new forces gain dynamical importance when the anisotropy $A = p_{\perp} / p_{\parallel}$ deviates significantly from unity (depending on $\beta$) and  give rise to instabilities. 
The standard CGL-MHD model is able to capture the correct linear behaviour of the long wavelength limit of the firehose instability (which has scales much larger than the proton Larmor radius $l_{cp}$), but not of the mirror instability which is overstable (see the kinetic and CGL-MHD instability limits in the $A-\beta_{\parallel}$ plane in Figure~\ref{fig:beta_vs_a}). The correct linear threshold of the mirror instability can be obtained from  higher order fluid models which evolve heat conduction (e.g.~\citealt{snyder_etal_1997, ramos_2003, kuznetsov_dzhalilov_2010}) and results in substantial difference with regard to the CGL-MHD criterium (see the kinetic and CGL-MHD instability limits in the $A-\beta_{\parallel}$ plane in Figure~\ref{fig:beta_vs_a}).

These same (mirror and firehose) instabilities are known to constrain the (proton) pressure anisotropy  growth to values close to the instability thresholds, via wave-particle interactions which obviously are not captured by any fluid model. Other kinetic instabilities driven by pressure/temperature anisotropy are also known to relax the anisotropy, such as the cyclotron instability (for protons) and whistler anisotropy instability (for electrons; see \citealt{gary_1993}).  
Based on this phenomenology, we here imposed source terms on the standard CGL-MHD equations which relaxed the pressure anisotropy $A$ to the marginally stable value (conserving the internal energy) at a rate $\nu_S$, whenever $A$ evolved to a value inside the unstable kinetic mirror or firehose zones. 

As remarked before, there are several studies about the rate at which instabilities driven by pressure anisotropy  relax the anisotropy itself. Using 2D particle simulations, \cite{gary_etal_2000} studied the anisotropy relaxation rate for protons subject to cyclotron instability and found rates which are related to the growth rate of the fastest unstable mode
 $\sim 10^{-3}-10^{-1} \Omega_p$ (where $\Omega_p$ is the proton gyrofrequency). \cite{nishimura_etal_2002}, also employing 2D particle simulations, found an analogous result for electrons subject to to the whistler anisotropy instability with an anisotropy relaxation rate of a few percent of the electron gyrofrequency. In both studies, part of the free energy of the instabilities is converted to magnetic energy.
Recently, \cite{yoon_seough_2012} and \cite{seough_yoon_2012} studied the saturation of specific modes of the mirror and firehose instabilities via quasi-linear calculations, using the Vlasov-Maxwell dispersion relation. They  also found that the temperature anisotropy relaxes to the marginal state after a few  hundreds of the proton Larmor period and there is  accumulation of magnetic fluctuations at the proton Larmor radius scales. 

However, exactly what kinetic instabilities saturate the pressure anisotropy or the detailed processes involved are not fully understood yet and one cannot be sure to what extent the rates inferred in the  studies above or those employed in the present analysis  are  applicable to the ICM plasma, specially with driven turbulence. In other words, the rate $\nu_S$  is subject to uncertainties and further forthcoming study involving particle-in-cell (PIC) simulations will be performed in order to investigate this issue in depth. 
In particular, in a very recent study about accretion disks, \cite{riquelme_etal_2012} performed direct two-dimensional PIC  shearing box simulations  and found that for low beta values ($\beta<0.3$), the  pressure anisotropy is constrained by the ion-cyclotron instability threshold, while for large beta values  the mirror instability threshold constrains  the anisotropy, which is compatible with the present study. However, they have also  found that in the low beta regime, initially  the  anisotropy can reach maximum values above  the threshold due to the mirror instability. Nevertheless, they have attributed this behaviour to  the initial cyclotron frequency adopted for the particles which was small compared to the orbital frequency in order to save computation time. 

We must add yet that here we have taken into account  the isotropization feedback due to the firehose and mirror instabilities only,  neglecting, for instance, the ion-cyclotron instability because this is more probably to be  important in low $\beta_{\parallel}$ regimes, which is not the case for ICM plasmas.
 We have considered that the anisotropy relaxation to the marginally stable value occurs at the rate of the fastest mode of the triggered  instability (Equations~\ref{eqn:max_growth_rates}), which is of the order of the proton gyrofrequency. As discussed above, for the typical parameters of the ICM, these relaxation times correspond to time scales which are extremely short compared to the shortest dynamical times one can solve. This means that  the plasma at least at the macroscopic scales is essentially  always inside the stable region. This justifies why we adopted the simple approach of constraining  the anisotropy by the marginal values of the instabilities, similarly to  the $hardwall$ constraints  employed by \cite{sharma_etal_2006}. However, if one could  resolve all the scales and frequencies of the system, one would probably detect some fraction of the plasma at the small scales lying in the unstable region. For the ICM, the scale of the fastest growing mode is $\sim 10^{10}$ cm, i.e., the proton Larmor radius.

\subsection{Consequences of assuming one-temperature approximation for all species}

Although  the electrons have a larger  collisional rate  than the protons in the ICM ($\sim \sqrt{m_p/m_e}$), we have assumed in this work, for simplicity, that both species have the same anisotropy in pressure. Also, we assumed them to be in ``thermal equilibrium''. A more precise approximation would be to consider the electrons  only with an isotropic pressure. This would require another  equation to evolve  the electronic pressure and additional physical ingredients in our model, such as a prescription on how to share the turbulent energy converted into heat at the end of the turbulent cascade or how to quantify the thermalization of the free-energy released by the kinetic instabilities, as well as a description of the cooling for each of the species. 
The assumption of same temperature and pressure anisotropy for both species has resulted a force on the collisionless plasma due to the latter which is maximized. Nevertheless, since our results have shown  that the dynamics of the turbulence when considering the relaxation of the  anisotropy due to the instabilities feedback is similar to that of collisional MHD, we can conclude that if we had considered  the electronic pressure to be already isotropic then,  this similarity would be even greater.

Another relevant aspect that should  be considered in future work regards the fact that  the electron thermal speed achieves relativistic values for  temperatures $\sim 10$ keV which are  typical in the ICM. Thus a  more consistent calculation would require a relativistic treatment (see for example \citealt{hazeltine_mahajan_2002}).

\subsection{Limitations of the thermal relaxation model}

Our model considers a  thermal relaxation (Eq.~\ref{eqn:thermal_relax}) which  ensures that the average temperature of the domain is maintained nearly constant, despite of the continuous  dissipation of turbulent power. This simplification allowed us to avoid a detailed  description of the radiative cooling and its influence on the temperature anisotropy. Even though, the rate $\nu_{th}=5$ employed  in most of our simulations  is low enough to not perturb significantly the dispersion relation arising from the CGL-MHD equations. Time scales $\delta t \simeq \nu_{th}^{-1} = 0.2$  are much larger than the typical time-step of our simulations ($\sim 10^{-5}$). This means that  the maximum characteristic speeds calculated via relations \eqref{eqn:cgl_alfven_speed} and \eqref{eqn:cgl_fastslow_speed} were more than appropriate  for the calculation of the  fluxes in our numerical scheme (see \S 3.1). 

In order to evaluate  the effects of the rate of the thermal relaxation on the turbulence statistics, we also performed numerical simulations of  two models with different rates $\nu_{th}$ (namely, models A5 and A6 of Table~\ref{tab:models}). Model A5 has a slower rate  than model A1, $\nu_{th}=0.5$, and the system suffers continuous increase of the temperature as time evolves which  increases  $\beta$  and reduces the sonic Mach number of the turbulence. Model A6, on the other hand, has a faster rate $\nu_{th}=50$ and quickly converges  to the isothermal limit. Despite of different averages and standard deviations in their internal energy, models A5 and A6 presented overall behaviour similar to  model A1 (see Table~\ref{tab:statistics_A}). 

We have also tested models  without anisotropy relaxation (not shown here) which employed the CGL-MHD  equations of state for calculating the pressure components parallel and perpendicular to the magnetic field (Equations~\ref{eqn:cgl_closure} accompanied of  homogeneous initial conditions, rather than  evolving  the two last equations in Eq.~\ref{eqn:collisionless_mhd} for the anisotropy $A$ and the internal energy, respectively). Although there are some intrinsic differences due to larger local values of the sound speeds, the overall behaviour of these models was qualitatively similar to the models with $\nu_S=0$ presented here.

In spite of the results above, a more accurate treatment of the energy evolution will be desirable in future work. For instance, as discussed earlier, the lack of a proper treatment for the heat conduction makes the linear behaviour of the mirror instability in a fluid description  different from the kinetic theory leading to  an overstability of the system. A higher order fluid model reproducing the kinetic linear behaviour of the mirror instability (see Eq. 9) would enhance the effects of this instability in the models with finite $\nu_S$, probably producing more small scale fluctuations compared to the present results (see Figure 5). 
The effects of the mirror instability on the turbulence statistics have been extensively discussed in \cite{kowal_etal_2011a} (see also next section) where a double-isothermal closure was used. This closure is able to reproduce the threshold of the mirror instability given by kinetic derivation.

\subsection{Comparison with previous studies}

\cite{kowal_etal_2011a} studied the statistics of the turbulence in  collisionless MHD flows assuming  fixed parallel and perpendicular temperatures in the so called double-isothermal approximation, but without taking into account the effects of anisotropy saturation due to the instabilities feedback. They explored  different regimes of turbulence (considering different combinations of sonic and Alfv\'enic Mach numbers) and initially different  (firehose or mirror) unstable regimes. They analysed the power spectra of the density and velocity, and also the anisotropy of the  structure function of these quantities and found that super-Alfv\'enic, supersonic turbulence in these double-isothermal collisionless models do not evidence  significant differences compared to the collisional-MHD counterpart. 

In the case of subsonic models, they have also detected  an increase in the density and velocity power spectra  
at the smallest scales due to the growth of the instabilities at these scales, when compared to the collisional-MHD counterparts. They found elongation of the density and velocity structures along the magnetic field in mirror unstable simulations and isotropization of these structures in the firehose unstable models. In the present study, the closest to  their models is the high $\beta$, super-Alfv\'enic model A2 which is without anisotropy relaxation. As in their subsonic sub-Alfv\'enic mirror unstable case, the instabilities accumulate power in the smallest scales of the density and velocity spectra. However, we should note that the density and velocity structures in our Model A2 become more isotropic at these scales probably  because it is in a super-Alfv\'enic regime.

Our simulations starting with initial seed magnetic field have revealed  the crucial role of the pressure anisotropy saturation (due to the mirror instability) for the dynamo turbulent amplification of the magnetic field, which in turn increases the anisotropy  $A$. In our seed field  simulations  without  anisotropy constraints (models C2 and C3), where the mirror forces dominate the dynamics, the turbulent flow is not able to stretch the field lines  and therefore, there is no magnetic field amplification. On the other hand, in  model C1 where the pressure anisotropy growth is constrained by the instabilities, there is a dynamo amplification  of  the magnetic energy until nearly equipartition with the kinetic energy. 
This result is in agreement with 3D numerical simulations of magneto-rotational instability (MRI) turbulence performed by \cite{sharma_etal_2006}, where a collisionless fluid model taking into account the effects of heat conduction was employed in a  shearing box.  They have found that  the anisotropic stress stabilizes the MRI when no bounds on the anisotropy are considered, making the magnetic lines stiff and  avoiding its amplification. When using bounds on the anisotropy, however, they found that the MRI generated is similar (but with some small quantitative corrections) to the collisional-MHD case.
\cite{sharma_etal_2006}, however,  did not consider any cooling mechanism, so that  the temperature  increased continuously  in  their simulations. Besides, the simulations here presented have substantially larger resolution.
Further, they have found  that the system overall evolution is nearly insensitive to the adopted  thresholds values for the anisotropy. We have also found little difference in the turbulence statistics between models with different non null values of the anisotropy relaxation rate.

\cite{meng_etal_2012b} also employed  a collisionless MHD model to investigate the Earth's magnetosphere by means of  3D global simulations. They employed the CGL closure, adding   terms to constrain the anisotropy in the ion pressure only (the electronic pressure considered isotropic was neglected in their study). Using  real data from the solar wind at the inflow boundary, they compared the outcome of the model in trajectories where data from space crafts (correlated to the inflow data) were available.
Then, they repeated the same calculation, but employing   a  collisional MHD model.  They found better agreement with the collisionless MHD model in the trajectory passing by the bowshock region, where gas is compressed in the direction parallel to the radial magnetic field lines, producing a firehose ($A<1$) unstable zone. However, in the trajectory passing by the magnetotail, the simulated data in the collisionless model were not found to be  more precise than in the collisional case. In summary, the collisionless MHD description of the magnetosheath seems to differ little from the standard MHD model when the anisotropy is constrained. Even though, they have found quantitative differences in, for example, the thickness of the magnetosheath, which  is augmented in the collisionless case, in better agreement with the observations. 
In the  more homogeneous problem discussed here, in a domain with periodic boundaries and isotropic turbulence driving, we have found that  both the evolution of the turbulence  and  the turbulent  dynamo growth  in the ICM under a collisionless-MHD description accounting  for the anisotropy saturation due to the kinetic instabilities feedback, behave similarly (both qualitatively and quantitatively)  to  the collisional-MHD description.
\footnote{We note, as remarked before, that while in the case of the ICM plasma the anisotropy relaxation rate is expected to be much larger than the dynamical rates of turbulent motions by several orders of magnitude, in the case of the solar wind the relaxation rate is only about ten times larger than the characteristic compression rates, so that in this case an instantaneous  relaxation of the anisotropy is not always applicable (Meng et al. 2012b, Chandran et al. 2011).  }

\cite{brunetti_lazarian_2011} appealed to theoretical arguments about the decrease of the effective
mean free path and related isotropization of the particle distribution to argue that the collisionless
damping of compressible modes will be reduced in the  ICM compared to the calculations in  earlier papers
 (\citealt{brunetti_lazarian_2007})\footnote{This happened to be important for cosmic ray acceleration by fast modes (see \citealt{yan_lazarian_2002, yan_lazarian_2004, yan_lazarian_2008}) that takes place in the ICM.}. Our
present calculations do not account for the collisionless damping of compressible motions, but similar to
\cite{brunetti_lazarian_2011} we may argue that this type of damping is not important at least for the
large scale compressions.

\subsection{Implications of the present study}

The dynamics of the ICM plasmas is important for understanding most of the ICM physics, including the formation of galaxy clusters and their evolution. The relaxation that we discussed in this paper explains how
clusters can have magnetic field generation, as well as turbulent cascade present there. We showed that
for sufficiently high rates of isotropization arising from the interaction of particles with magnetic
fluctuations induced by plasma instabilities, the collisionless plasma becomes {\it effectively collisional}
and can be described by ordinary MHD approach. This can serve as a justification for earlier MHD studies
of the ICM dynamics and can motivate new ones. 

In general,  ICM studies face one major problem. 
The estimated Reynolds number for the ICM using the Coulomb cross-sections is  small ($\sim 100$ or less) so that one may even question the existence of
turbulence in galaxy clusters. 
 This is the 
problem that we deal with in the present paper and argue that the  Reynolds numbers in the ICM may
be much larger than the naive estimates above. The difference comes from the dramatic decrease of the mean 
free path of the particles due to the interaction of ions with fluctuations induced by plasma instabilities.  
In other words, our study shows that the collisional MHD approach may correctly represent properties of turbulence in the intracluster plasma. In particular, it indicates that MHD turbulence theory 
may be applicable to a variety of collisionless media. 
This is a big extension of the
domain of applicability of the \cite{goldreich_sridhar_1995} theory of Alfv\'enic turbulence.

\section{Summary and Conclusions}

The plasma in the ICM is formally weakly collisional. Indeed, as far as Coulomb collisions are involved, 
the mean free paths of particles are comparable to size of galaxy clusters as a result of
 the high temperatures and low densities of the intracluster plasmas. Therefore, one might expect the plasmas
to have high viscosity and not allow turbulent motions. At the same time, magnetic fields and turbulence are observed to be present there. The partial resolution of the paradox may be that even small magnetic
fields can substantially decrease the perpendicular viscosity of plasmas and enable Alfv\'enic turbulence
that is weakly couple with the compressible modes (see also \citealt{lazarian_2006b}). Our present work indicates that
the parallel viscosity of plasmas can also be reduced compared with the standard Braginskii values.

 Aiming to understand the effects of the low collisionality on the turbulence statistics and on the turbulent magnetic field amplification in the ICM, both of which are  commonly treated using a collisional-MHD description, we performed three-dimensional numerical simulations of forced turbulence employing a single-fluid collisionless-MHD model. We focused on models with trans-sonic turbulence and at the high $\beta$ regime (where $\beta$ is the ratio between the thermal and magnetic pressures), which are conditions appropriate to the ICM. We also considered a model with low $\beta$ for comparison. 

Our collisionless-MHD approach is based on the CGL-MHD model, the simplest fluid model for a collisionless plasma, which differs from the standard collisional-MHD by the presence of an anisotropic thermal pressure tensor. The new forces arising from this anisotropic pressure modify the MHD linear waves and produce the firehose and mirror instabilities. These instabilities in a macroscopic fluid  can be viewed as the long wavelength limit of the corresponding kinetic instabilities driven by the temperature anisotropy for which the higher the $\beta$ regime the faster the growth rate.

Considering the feedback of the kinetic instabilities on the pressure anisotropy, we adopted a
plausible model of anisotropy relaxation and modified the CGL-MHD equations in order to take into account the effects of relaxation of the anisotropy arising from the scattering of individual ions by
fluctuations induced by plasma instabilities. This model appeals to  
 earlier observational and numerical studies in the context of the solar magnetosphere,
as well as theoretical considerations discussed in earlier works. While the details of this isotropization feedback are difficult to quantify from first principles, 
the rate at which an initial anisotropy is relaxed is found (at least in 2D PIC simulations) to be a few percent of the ion Larmor frequency (\citealt{gary_etal_1997, gary_etal_1998, gary_etal_2000}). The frequencies that  we deal with  in our numerical simulations are much larger than the ion Larmor frequency in the ICM (considering the scale of the computational domain $\sim$ $100$ kpc). This has motivativated  us to consider this anisotropy relaxation to be instantaneous. Nevertheless,   for completeness we also performed simulations with finite rates, in order to access their potential effects in   the results.

The main results from our simulated models can be summarized as follows:

\begin{itemize}

\item Anisotropy in the collisionless fluid is naturally created by turbulent motions as a consequence of fluctuations of the magnetic field and gas densities. In all our models, the net increase of magnetic field intensity led to the predominance of the perpendicular pressure in most of the volume of the domain;

\item In the high $\beta$ regime with moderate initial magnetic field, the model without anisotropy relaxation (which is therefore, a ``standard'' CGL-MHD model; see Model A2 in Figures~\ref{fig:maps_dens} and~\ref{fig:beta_vs_a}) has the PDF of the density broadened, specially in the low density tail, in comparison to the collisional-MHD model. 
This is a consequence of the action of the mirror instability which traps the gas in small cells of low magnetic field intensity. 
The density and velocity power spectra show excess of power specially at small scales, where the instabilities are stronger, although the magnetic field reveals less power. Consistently, the anisotropies in the structure functions of density, velocity, and magnetic field are reduced at the smallest scales in comparison to the collisional-MHD model;

\item Models with anisotropy relaxation (either instantaneous, or with the finite rates  $~10^{2}$ times or $10^{3}$ times larger than the inverse of the turbulence turnover time $t_{turb}$) present density PDFs, power spectra, and anisotropy in structures which are very similar to the collisional MHD model. However, the model with the smallest anisotropy relaxation rate ($\sim 10^{2}t_{turb}^{-1}$) shows a little excess of power in density and velocity in the smallest scales, already in the dissipative range. This is consistent with the presence of instabilities in the smallest regions of the gas;

\item Models starting with a very weak, seed magnetic field (i.e., with very high $\beta$), without any anisotropy relaxation, have the magnetic energy saturated at levels many orders of magnitude smaller than kinetic energy. The value of the magnetic energy at this saturated state is shown to depend on the sonic Mach number of the turbulence, the smaller the sound speed the higher this saturation value;

\item Models starting with a very weak, seed magnetic field, but with anisotropy relaxation (with instantaneous or finite  rates) show an increase of the magnetic energy until values close those achieved by the collisional-MHD model. The growth rate of the magnetic energy for the model with instantaneous relaxation rate is similar to the collisional-MHD model, but this rate is a little smaller for the models with a finite rate of the anisotropy relaxation, as one should expect;

\item In the low $\beta$ regime, the strength of the injected turbulence (trans-sonic and sub-Alfv\'enic) is not able to produce anisotropy fluctuations which trigger instabilities. The statistics of the turbulence is very similar to the collisional-MHD case, in consistency with the fact that in this regime the pressure forces have minor importance.

\end{itemize}

All these results show that the applicability of the collisional-MHD approach for studying the dynamics of the ICM, especially in the turbulent dynamo amplification of the magnetic fields, is justified if the anisotropy relaxation rate provided by the kinetic instabilities is fast enough and the anisotropies are relaxed until the marginally stable values. As stressed before, the quantitative description of this process is still lacking, but if we assume that the results obtained for the anisotropy relaxation (usually studied in the context of the collisionless plasma of the solar wind) can be applied to the turbulent ICM, we should expect a relaxation rate much faster than the rates at which the anisotropies are created by the turbulence. 

We intend in future work to investigate the kinetic instabilities feedback on the pressure anisotropy in the context of the turbulent ICM. To do this in a self-consistent way a kinetic approach is required. 
This can be done analytically (similar to the calculations in e.g. \citealt{yan_lazarian_2011} for collisionless
fluid of cosmic rays) and/or by the employment of PIC simulations. We feel that gauging  our approach
with PIC simulations is particularly straightforward and should be done first.

We should emphasize that, even in the case of a good agreement between the collisional-MHD and collisionless-MHD results for the dynamics of the ICM,  collisionless effects, like the kinetic instabilities themselves, can still be important for energetic processes in the ICM, such as  the acceleration of particles (Kowal et al. 2011b, 2012b) and heat conduction (\citealt{narayan_medvedev_2001, schekochihin_etal_2010, kunz_etal_2011, rosin_etal_2011, riquelme_etal_2012}).

\acknowledgments
RSL acknowledges support from a grant of the Brazilian Agency FAPESP (2007/04551-0), EMGDP from  FAPESP (2006/50654-3, and 2011/53275-4) and CNPq (306598/2009-4) grants, and  DFG from  CNPq (no. 300382/2008-1) and FAPESP (no. 2011/12909-8). AL acknowledges the NSF grant AST  1212096 and the NSF-sponsored Center for Magnetic Self-Organization, as well as the stimulating atmosphere of the International Institute  of 
Physics (Natal, Brazil) and of the Harvard University during his sabbatical.
The numerical simulations of this work were performed with the super computer of the Laboratory of Astro-informatics LAi (IAG/USP, NAT/Unicsul), whose purchase was made possible by  FAPESP (grant 2009/54006-4).

\begin{appendix}

\section{Supplementary tables and figures}

Tables~\ref{tab:statistics_A},~\ref{tab:statistics_B}, and~\ref{tab:statistics_C} present one point statistics in space and time for the simulated models in Table~\ref{tab:models}. The averaged quantities are listed in the most left column. Each column presents the averages and bellow it the standard deviation, for each model. 
For the statistics, we considered snapshots spaced in time by $\Delta t = 1$, from $t=2)$ (Tables~\ref{tab:statistics_A} and~\ref{tab:statistics_B}) or $t_{f}-10$ (Table~\ref{tab:statistics_C}), until $t_{f}$ (the $t_{f}$ for each model is listed in Table~\ref{tab:models}). All the values are in code units and can be converted into physical units according to the prescription given in \S 3.2. The functional definitions (in terms of the code units) of the physical quantities listed are: $E_K = \rho u^{2}/2$, $E_M = B^{2}/2$, $E_I = (p_{\perp}+p_{\parallel}/2)$, $M_A = u \rho^{1/2}/B$, $M_S = u (3 \rho)^{1/2} / (2 p_{\perp} + p_{\parallel})^{1/2}$. For the collisional MHD models, the following definitions are used: $E_I = 3p/2$, $M_S = u (\rho/p)^{1/2}$, $\beta_{\parallel} = \beta$.

\begin{table*}
\caption{Space and time averages (upper lines) and standard deviations (lower lines) for the models A which have moderate initial magnetic fields ($\beta_0= 200$).}
\centering
\begin{tabular}{l | c | c | c | c | c | c | c}
\hline \hline
Quantity &
A1 &
A2 &
A3 &
A4 &
A5 &
A6 &
Amhd \\
[0.5ex]
\hline
$\langle \log{\rho} \rangle$ &
$-6.3 \times 10^{-3}$ &
$-2.7 \times 10^{-2}$ &
$-6.3 \times 10^{-3}$ &
$-5.9 \times 10^{-3}$ &
$-2.6 \times 10^{-3}$ &
$-1.0 \times 10^{-2}$ &
$-8.0 \times 10^{-3}$
\\
&
$7.5 \times 10^{-2}$ &
$0.17$ &
$7.5 \times 10^{-2}$ &
$7.3 \times 10^{-2}$ &
$4.8 \times 10^{-2}$ &
$9.5 \times 10^{-2}$ &
$8.5 \times 10^{-2}$
\\
\hline
$\langle u^2 \rangle$ &
$0.48$ &
$0.57$ &
$0.46$ &
$0.47$ &
$0.48$ &
$0.50$ &
$0.55$
\\
&
$0.40$ &
$0.52$ &
$0.39$ &
$0.39$ &
$0.41$ &
$0.41$ &
$0.48$
\\
\hline
$\langle E_K \rangle$ &
$0.24$ &
$0.27$ &
$0.23$ &
$0.23$ &
$0.23$ &
$0.24$ &
$0.27$
\\
&
$0.20$ &
$0.24$ &
$0.19$ &
$0.19$ &
$0.20$ &
$0.21$ &
$0.24$
\\
\hline
$\langle E_M \rangle$ &
$0.25$ &
$0.12$ &
$0.24$ &
$0.25$ &
$0.25$ &
$0.24$ &
$0.29$
\\
&
$0.16$ &
$0.18$ &
$0.16$ &
$0.16$ &
$0.17$ &
$0.15$ &
$0.23$
\\
\hline
$\langle E_I \rangle$ &
$1.7$ &
$1.7$ &
$1.7$ &
$1.6$ &
$2.9$ &
$1.5$ &
$1.7$
\\
&
$0.39$ &
$0.54$ &
$0.39$ &
$0.38$ &
$0.56$ &
$0.34$ &
$0.44$
\\
\hline
$\langle M_A \rangle$ &
$1.2$ &
$1.7$ &
$1.2$ &
$1.2$ &
$1.2$ &
$1.2$ &
$1.3$
\\
&
$1.5$ &
$1.2$ &
$1.6$ &
$1.5$ &
$1.6$ &
$1.5$ &
$1.5$
\\
\hline
$\langle M_S \rangle$ &
$0.60$ &
$0.67$ &
$0.59$ &
$0.60$ &
$0.45$ &
$0.64$ &
$0.64$
\\
&
$0.26$ &
$0.31$ &
$0.26$ &
$0.25$ &
$0.20$ &
$0.28$ &
$0.29$
\\
\hline
$\langle \log{A} \rangle$ &
$1.8 \times 10^{-2}$ &
$0.45$ &
$2.0 \times 10^{-2}$ &
$1.6 \times 10^{-2}$ &
$6.0 \times 10^{-3}$ &
$1.9 \times 10^{-2}$ &
-
\\
&
$9.0 \times 10^{-2}$ &
$0.65$ &
$0.10$ &
$9.1 \times 10^{-2}$ &
$5.8 \times 10^{-2}$ &
$9.5 \times 10^{-2}$ &
-
\\
\hline
$\langle \log{\beta_{\parallel}} \rangle$ &
$0.74$ &
$0.70$ &
$0.77$ &
$0.75$ &
$1.0$ &
$0.71$ &
$0.75$
\\
&
$0.47$ &
$1.0$ &
$0.49$ &
$0.47$ &
$0.47$ &
$0.46$ &
$0.52$
\\
\hline
$\langle | p_{\parallel} - p_{\perp} | / \left( B^{2} / 4 \pi \right) \rangle$ &
$0.37$ &
$4.9$ &
$3.6$ &
$1.4$ &
$0.41$ &
$0.37$ &
-
\\
&
$0.25$ &
$3.2 \times 10^{2}$ &
$1.0 \times 10^{3}$ &
$8.5 \times 10^{2}$ &
$0.26$ &
$0.25$ &
-
\\
[1ex]
\hline
\end{tabular}
\label{tab:statistics_A}
\end{table*}

\begin{table*}
\caption{Space and time averages (upper lines) and standard deviations (lower lines) for models B which have initial strong magnetic field ($\beta=0.2$).}
\centering
\begin{tabular}{l | c | c}
\hline \hline
Quantity &
B1 &
Bmhd \\
[0.5ex]
\hline
$\langle \log{\rho} \rangle$ &
$-1.0 \times 10^{-2}$ &
$-1.8 \times 10^{-2}$
\\
&
$9.8 \times 10^{-2}$ &
$0.12$ 
\\
\hline
$\langle u^2 \rangle$ &
$0.90$ &
$0.86$ 
\\
&
$0.79$ &
$0.73$
\\
\hline
$\langle E_K \rangle$ &
$0.44$ &
$0.42$ 
\\
&
$0.41$ &
$0.39$ 
\\
\hline
$\langle E_M \rangle$ &
$4.8$ &
$4.8$ 
\\
&
$0.80$ &
$0.90$ 
\\
\hline
$\langle E_I \rangle$ &
$1.7$ &
$1.7$ 
\\
&
$0.57$ &
$0.75$ 
\\
\hline
$\langle M_A \rangle$ &
$0.28$ &
$0.27$ 
\\
&
$0.13$ &
$0.13$ 
\\
\hline
$\langle M_S \rangle$ &
$0.83$ &
$0.82$ 
\\
&
$0.37$ &
$0.36$ 
\\
\hline
$\langle \log{A} \rangle$ &
$5.5 \times 10^{-2}$ &
-
\\
&
$0.21$ &
-
\\
\hline
$\langle \log{\beta_{\parallel}} \rangle$ &
$-0.69$ &
$-0.66$
\\
&
$0.30$ &
$0.21$
\\
\hline
$\langle | p_{\parallel} - p_{\perp} | / \left( B^{2} / 4 \pi \right) \rangle$ &
$4.1 \times 10^{-2}$ &
-
\\
&
$3.9 \times 10^{-2}$ &
-
\\
[1ex]
\hline
\end{tabular}
\label{tab:statistics_B}
\end{table*}

\begin{table*}
\caption{Space and time averages (upper lines) and standard deviations (lower lines) for models C which have initial very weak (seed) magnetic field.}
\centering
\begin{tabular}{l | c | c | c | c | c }
\hline \hline
Quantity &
C1 &
C2 &
C3 &
C4 &
Cmhd \\
[0.5ex]
\hline
$\langle \log{\rho} \rangle$ &
$-8.5 \times 10^{-3}$ &
$-1.5 \times 10^{-2}$ &
$-8.7 \times 10^{-2}$ &
$-8.8 \times 10^{-3}$ &
$-9.1 \times 10^{-3}$
\\
&
$8.7 \times 10^{-2}$ &
$0.12$ &
$0.28$ &
$8.9 \times 10^{-2}$ &
$9.0 \times 10^{-2}$
\\
\hline
$\langle u^2 \rangle$ &
$0.79$ &
$0.70$ &
$0.79$ &
$0.80$ &
$0.79$
\\
&
$0.63$ &
$0.73$ &
$0.64$ &
$0.63$ &
$0.63$
\\
\hline
$\langle E_K \rangle$ &
$0.38$ &
$0.32$ &
$0.37$ &
$0.38$ &
$0.38$
\\
&
$0.30$ &
$0.31$ &
$0.39$ &
$0.29$ &
$0.30$
\\
\hline
$\langle E_M \rangle$ &
$6.2 \times 10^{-2}$ &
$6.2 \times 10^{-6}$ &
$1.0 \times 10^{-4}$ &
$2.6 \times 10^{-2}$ &
$9.0 \times 10^{-2}$ 
\\
&
$7.5 \times 10^{-2}$ &
$3.5 \times 10^{-5}$ &
$3.7 \times 10^{-4}$ &
$3.9 \times 10^{-2}$ &
$0.11$ 
\\
\hline
$\langle E_I \rangle$ &
$1.7$ &
$1.7$ &
$0.33$ &
$1.6$ &
$1.7$
\\
&
$0.49$ &
$0.44$ &
$0.28$ &
$0.49$ &
$0.51$
\\
\hline
$\langle M_A \rangle$ &
$4.6$ &
$5.9 \times 10^{2}$ &
$2.8 \times 10^{2}$ &
$8.1$ &
$3.8$
\\
&
$6.7$ &
$5.7 \times 10^{2}$ &
$4.2 \times 10^{2}$ &
$11$ &
$5.3$ 
\\
\hline
$\langle M_S \rangle$ &
$0.78$ &
$0.74$ &
$2.1$ &
$0.79$ &
$0.78$
\\
&
$0.34$ &
$0.38$ &
$1.0$ &
$0.34$ &
$0.34$
\\
\hline
$\langle \log{A} \rangle$ &
$1.3 \times 10^{-2}$ &
$0.51$ &
$2.1$ &
$6.7 \times 10^{-3}$ &
-
\\
&
$3.2 \times 10^{-2}$ &
$0.87$ &
$1.2$ &
$8.6 \times 10^{-2}$ &
-
\\
\hline
$\langle \log{\beta_{\parallel}} \rangle$ &
$1.5$ &
$5.5$ &
$2.0$ &
$2.0$ &
$1.4$
\\
&
$0.64$ &
$1.4$ &
$2.1$ &
$0.70$ &
$0.62$
\\
\hline
$\langle | p_{\parallel} - p_{\perp} | / \left( B^{2} / 4 \pi \right) \rangle$ &
$0.52$ &
$8.8 \times 10^{5}$ &
$5.5 \times 10^{4}$ &
$1.5 \times 10^{2}$ &
-
\\
&
$0.25$ &
$7.1 \times 10^{7}$ &
$1.7 \times 10^{7}$ &
$6.7 \times 10^{4}$ &
-
\\
[1ex]
\hline
\end{tabular}
\label{tab:statistics_C}
\end{table*}

\begin{figure*}
\centering
\begin{tabular}{c c}
\input{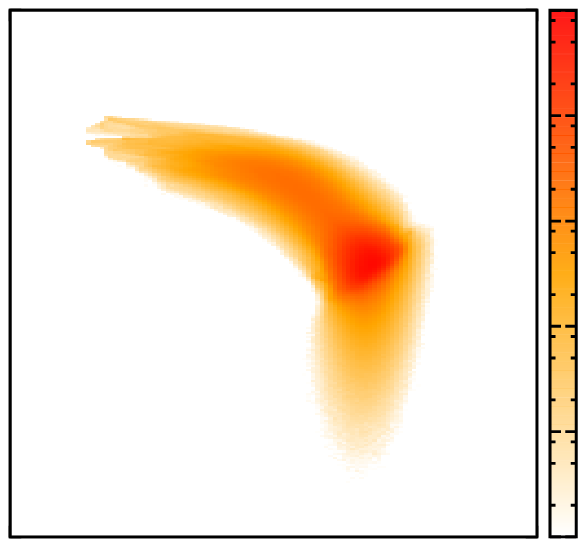} &
\input{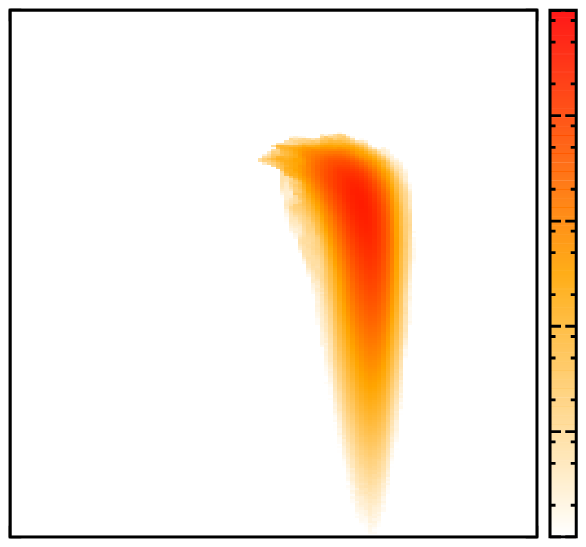} 
\end{tabular}
\caption{Two-dimensional normalized histograms of $\log{\rho}$ versus $\log{B}$. Left: collisionless model A2 with null anisotropy relaxation rate. Right: collisional MHD model Amhd. The histograms were calculated using snapshots every $\Delta t = 1$, from $t=2$ until the final time $t_{f}$ indicated in Table~\ref{tab:models}. See more details in \S 4.3.}
\label{fig:pdf_ldn_lb}
\end{figure*}

\begin{figure*}
\centering
\begin{tabular}{c c}
\input{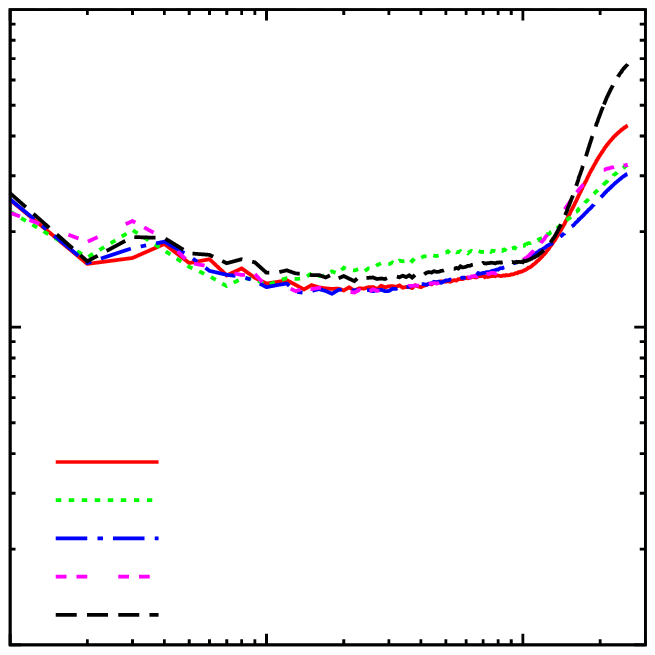} &
\input{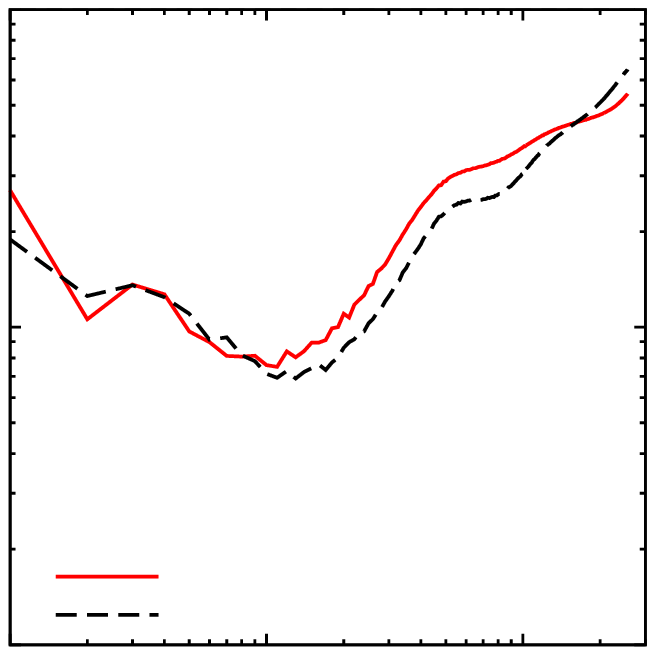}
\end{tabular}
\caption{Ratio between the power spectrum of the compressible component ($P_C(k)$) and the total velocity field ($P_u(k)$), for the same models as in Figure~\ref{fig:ps_turb} (see  Table~\ref{tab:models} and \S 4.4 for details).}
\label{fig:ps_velcvelo}
\end{figure*}

\begin{figure*}
\centering
\begin{tabular}{c c}
\input{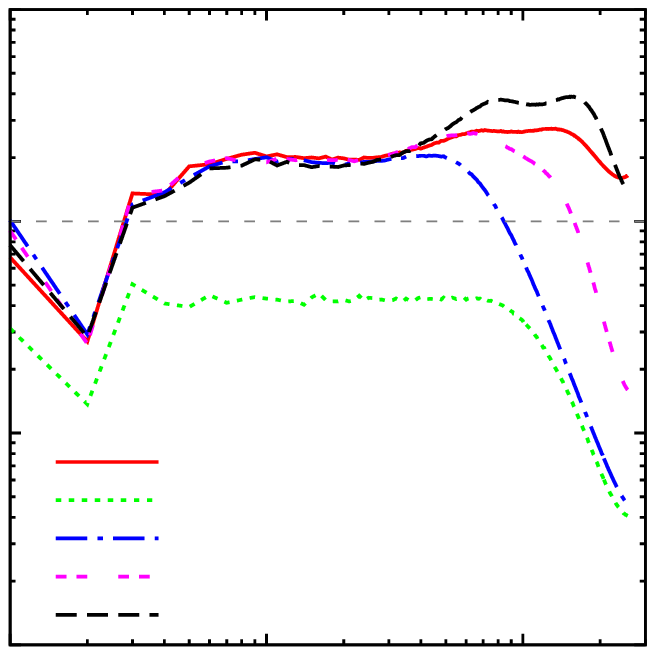} &
\input{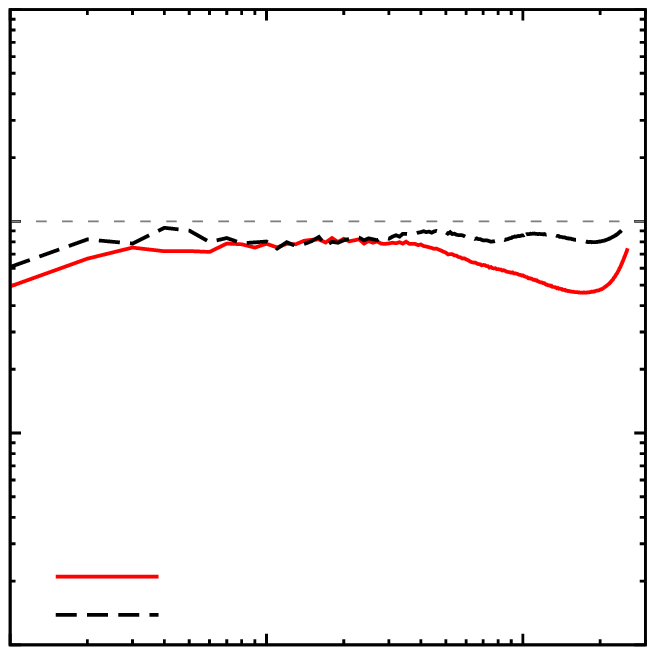}
\end{tabular}
\caption{Ratio between the power spectrum of the magnetic field ($P_B(k)$) and the velocity field ($P_u(k)$) for the same models as in Figure~\ref{fig:ps_turb} (see  Table~\ref{tab:models}
and \S 4.4 for details).}
\label{fig:ps_magnvelo}
\end{figure*}

\end{appendix}

\end{document}